\documentclass[12pt,a4paper]{article}
\usepackage[english]{babel}
\usepackage[utf8]{inputenc}
\usepackage[T1]{fontenc}
\usepackage{amsmath}
\usepackage{amsfonts}
\usepackage{amssymb}
\usepackage[colorinlistoftodos, color=green!40, prependcaption]{todonotes}
\usepackage{authblk}
\usepackage{cite}
\usepackage{amsthm}
\usepackage{mathtools}
\usepackage{xcolor}
\usepackage{graphicx}
\usepackage{adjustbox}
\usepackage{placeins}
\usepackage[T1]{fontenc}
\usepackage{lipsum}
\usepackage{csquotes}

\usepackage[linktoc=page]{hyperref}
\hypersetup{
    pdfauthor={F. Ambrosino, A. Duci, P. A. Grassi, and S. Penati},
    colorlinks=true, citecolor=blue, urlcolor=blue,
    pdfstartview={FitH}, 
pdftitle={SuSymTFT}, 
linkcolor=blue, 
citecolor=blue, 
filecolor=magenta, 
}

\usepackage{pifont}

\usepackage[top=2cm,bottom=2.5cm,outer=2.5cm,inner=2.5cm,marginparwidth=2.5cm]{geometry}
\usepackage{authblk} 
\usepackage{mathrsfs}
\usepackage{xcolor}
\usepackage{setspace}
\usepackage{placeins}
\newcommand{\Res}[1]{\mathop{\mathrm{Res}}_{#1}}

\newcommand\frontmatter{%
\clearpage
\pagenumbering{roman}
}

\newcommand\mainmatter{%
\clearpage
\pagenumbering{arabic}
}

\usepackage{amsmath}
\usepackage{amsfonts}
\usepackage{amssymb}

\usepackage{lmodern} 
\usepackage{caption}
\usepackage{subcaption} 
\usepackage{titlesec} 
\usepackage{fancyhdr} 
\usepackage{abstract} 
\usepackage{tocloft} 
\usepackage{comment}

\usepackage{hyperref}
\hypersetup{
pdfstartview={FitH},
pdftitle={Meromorphic amplitudes from 3d supersymmetry},
pdfauthor={F. Ambrosino, N. Haouzi},
pdfsubject={},
pdfcreator={},
pdfkeywords={},
colorlinks=true,
citecolor=blue,
linkcolor=blue,
urlcolor=blue
}

\numberwithin{equation}{section}
\usepackage{mathtools}
\usepackage{tikz}
\usepackage{tikz-cd}

\newcommand{\beq}{\begin{equation}}
\newcommand{\eeq}{\end{equation}}
\newcommand{\q}{\mathfrak q}
\newcommand{\ft}{\mathfrak t}
\newcommand{\ff}{\mathfrak f}
\newcommand{\cN}{\mathcal{N}}
\newcommand{\cT}{\mathcal{T}}
\newcommand{\cB}{\mathcal{B}}
\newcommand{\cD}{\mathcal{D}}
\newcommand{\cZ}{\mathcal{Z}}

\newcommand{\bZ}{\mathbb{Z}}
\newcommand{\bR}{\mathbb{R}}
\newcommand{\bC}{\mathbb C}
\newcommand{\dd}{\mathrm d}

\newcommand{\xmark}{\ding{55}}

\title{\Huge Meromorphic amplitudes from 3-dimensional supersymmetry 
\vspace{5mm}}

\author{Federico~Ambrosino%
\thanks{\href{mailto:federicoambrosino25@gmail.com}{federicoambrosino25@gmail.com}}, and Nathan Haouzi%
\thanks{\href{mailto:nathanhaouzi@gmail.com}{nathanhaouzi@gmail.com}}}
\affil{\normalsize \textit{Perimeter Institute for Theoretical Physics, Waterloo, Ontario N2L 2Y5, Canada}}

\date{\today}

\newcommand{\FA}[1]{\textcolor{olive}{[FA: #1]}}

\begin{document}

\frontmatter
\makeatletter\let\oldfnsymbol\@fnsymbol
\renewcommand{\@fnsymbol}[1]{%
  \ifcase#1
  \or $\q$%
  \or $\widetilde{\q}    $%
  \else \@arabic{#1}%
  \fi
}
\makeatother
\maketitle
\begin{abstract} 
We establish a new connection between supersymmetric  theories and scattering amplitudes. We show that the Coon amplitude coincides with the 3d $\mathcal{N}=2$ half-index of the XYZ model with nontrivial boundary conditions.  Our 3d theory, intrinsically defined in the UV, flows  to a sigma model  in the IR whose partition function is the Veneziano amplitude. Crossing symmetry is realized as a consequence of 3d $\cN=2$ mirror symmetry between XYZ and SQED. 
We use this correspondence to construct a meromorphic modification of the Coon amplitude by  promoting the long-standing dressing factor $\q^{ST}$  responsible for a branch cut to an elliptic completion thereof. This illustrates that one does not have to give up single-valuedness to achieve positivity at the physical poles.
\end{abstract}

\thispagestyle{empty}

\mainmatter

\tableofcontents
\newpage
\section{Introduction}
In the seminal paper \cite{Veneziano:1968yb} Veneziano proposed a model for a meromorphic, crossing symmetric, and unitary amplitude that exhibits a Regge behavior, now famously referred to as the Veneziano amplitude: \begin{equation}\label{Veneziano}
    A^{\rm Ven}(\alpha(s),\alpha(t)) =
\frac{\Gamma(-\alpha(s))\,\Gamma(-\alpha(t))}{\Gamma(-\alpha(s)-\alpha(t))}, \qquad \alpha(x) = m^2 + \alpha' x\, .
\end{equation}
This amplitude marked the dawn of string theory, and has since then been a lamppost to understand the structure of meromorphic amplitudes.
Soon thereafter \cite{Coon:1969yw}, Coon proposed a 1-parameter generalization of the Veneziano amplitude:
\begin{equation}\label{bareCoon}
A^{\text{bare}}_\q(\alpha_\q(s),\alpha_\q(t))=
\frac{\Gamma_\q(-\alpha_\q(s))\,\Gamma_\q(-\alpha_\q(t))}{\Gamma_\q(-\alpha_\q(s)- \alpha_\q(t))}\, .
\end{equation}
This is a $\q$-analog of the Beta function 
with very appealing physical properties, reducing to the Veneziano amplitude as $\q\to 1$. Exactly as the Veneziano amplitude, it is a crossing symmetric meromorphic function of $(s,t)$ with simple poles. 
This modification of the Veneziano amplitude has a long history\cite{Coon:1969yw,Baker:1970vxk,Coon:1972qz, Romans:1988qs, 
Romans:1989di,GonzalezMestres:1975ord}, but only in recent years, starting from \cite{Caron-Huot:2016icg}, there has been a surge in interest in Coon and other Veneziano deformations \cite{Figueroa:2022onw,Geiser:2022icl, Chakravarty:2022vrp, Geiser:2023qqq, Bhardwaj:2022lbz,
Maldacena:2022ckr, 
 Jepsen:2023sia, Cheung:2023adk, Wang:2024wcc, Cheung:2025tbr ,Komatsu:2025onf}.  
In physical Mandelstam variables, along the logarithmic trajectories
\begin{equation}\begin{aligned}
\sigma(s)&=1+(s-m^2)(\q-1),
\qquad &\tau(t)&=1+(t-m^2)(\q-1), \\
S=\alpha_\q(s)&:=\frac{\log\sigma(s)}{\log \q},
\qquad &T=\alpha_\q(t)&:=\frac{\log\tau(t)}{\log \q}\, ,
\end{aligned}
\end{equation}
there is an accumulation of poles in the spectrum at the finite point $s_\infty$:
\begin{equation}
  s_n=m^2+\frac{1-\q^n}{1-\q}=m^2+[n]_\q,
  \qquad
  [n]_\q:=\frac{1-\q^n}{1-\q}, \qquad   s_\infty
  =
  m^2+\frac{1}{1-\q}.
\end{equation}
This is the characteristic feature of the Coon amplitude and makes it an interesting toy model for meromorphic amplitudes with resonances located at finite distance in the $s$-plane.

\subsection[Amplitudes from 3d \texorpdfstring{$\mathcal{N}=2$}{N=2} theories]{\boldmath Amplitudes from 3d \texorpdfstring{$\mathcal{N}=2$}{N=2} theories}
The idea that string amplitudes have a supersymmetric interpretation is not new: consider a 2d $\cN=(2,2)$ abelian gauged linear sigma model, i.e.\ a $U(1)$ gauge theory with two charged chiral multiplets in the UV. For a positive choice of Fayet-Iliopoulos parameter, the theory flows in the IR to the $\mathbb{CP}^1$ nonlinear sigma model on the Higgs branch, and the Veneziano amplitude \eqref{Veneziano} can be written in integral form as:
\begin{equation}\label{period}
\begin{aligned}
    A^{\rm Ven}(\alpha(s),\alpha(t)) &=
    \int_0^1 dz \;z^{-\alpha(s)-1}\, (1-z)^{-\alpha(t)-1}  :=\int_{\Gamma}dz\; e^{-\widetilde{W}(z)}\, ,\\
    \widetilde{W}(z) &:=(1+\alpha(s)) \log(z) + (1+\alpha(t)) \log(1-z)\, .
    \end{aligned}
\end{equation}
The period integral on the right-hand side is naturally interpreted as the central charge of a Lagrangian brane, where $z$ is a local affine coordinate on a $\mathbb{CP}^1$ patch, 
 $\widetilde{W}(z)$ is a multivalued twisted superpotential encoding monodromies of a rank-1 local system on $\mathbb{CP}^1\backslash \{0,1,\infty\}$ via the Mandelstam variables, and the contour $\Gamma=[0,1]$ is a Lefschetz thimble supporting the brane for that potential \cite{aomoto1975equations,gelfand1986general,gelfand1989hypergeometric,hori2000mirror,hori2013exact}.
 
Here, we  show that the Coon amplitude also has a supersymmetric interpretation, but this time as the (half)-index of a 3-dimensional $\cN=2$ theory, with nontrivial boundary conditions (b.c.) . One essential feature is that while the Veneziano amplitude is intrinsically a sigma model quantity defined in the IR, the Coon amplitude index is instead inherently defined in the UV, meaning it can be computed straight from the classical fields of a 3d Lagrangian. In this sense, Coon should be regarded as a ``UV completion'' of the Veneziano amplitude.

Up to trivial normalization, we find that the Coon amplitude is the index of \textbf{the XYZ model}, a theory of 3 chiral multiplets $\rm X$, $\rm Y$, $\rm Z$, subjected to a tree-level superpotential $W=\rm XYZ$, with a mix of standard Neumann and Dirichlet b.c.\ . 
In fact, there is not just one but several 3d theories (with different choices of b.c.) which realize the Coon amplitude, a direct consequence of \textbf{3d mirror symmetry}. Namely, the residue expansion of the Coon amplitude in the $S$ (or $T$) channel should be understood in a mirror frame, as the index of 3d $\cN=2$ SQED. The latter is a $U(1)$ gauge theory with two charged chiral multiplets, whose global symmetries are an axial $U(1)_a$ flavor group and a $U(1)_{\rm top}$ topological symmetry. In this gauge theory, we find that the two Mandelstam variables $S$ and $T$ of the Coon amplitude are realized as the $U(1)_{\rm top}$ \textbf{Fayet-Iliopoulos (F.I.) parameter} $\xi$ and the $U(1)_a$ \textbf{axial mass} $a$, respectively:

\vspace{-.7cm}
$$\centering\includegraphics[width = 0.5\textwidth]{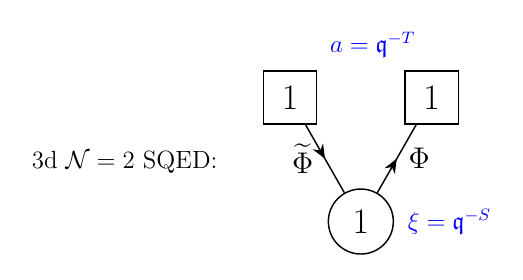}\label{quiverSQED}$$

\vspace{-0.5cm}
Remarkably, the crossing symmetry between the $S$ and $T$ channels of the amplitude is realized as a ``self-duality'' of the SQED gauge theory 
under the exchange of axial and topological symmetries $U(1)_a\leftrightarrow U(1)_{\rm top}$, which is manifest given our choice of boundary conditions: 
$$
\includegraphics[width =0.8\textwidth]{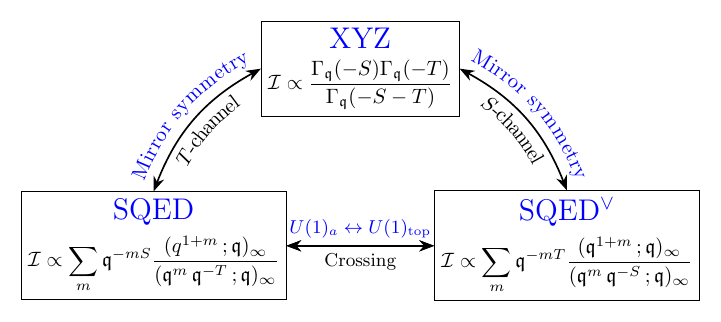}
$$
This illustrates a deep interplay between dualities and kinematical properties of the amplitude. From this perspective,  mirror symmetry gives a microscopic interpretation of crossing symmetry of the Veneziano amplitude after taking the $\q\to 1$ limit.

The precise b.c.\ needed in the SQED theory goes by the name of \textbf{Exceptional Dirichlet} \cite{Bullimore:2016nji}, and has attracted a lot of attention lately in mathematical physics \cite{Bullimore:2020jdq,Okazaki:2020lfy,Dedushenko:2021mds,Bullimore:2021rnr,Aganagic:2017smx,Haouzi:2023doo}.  
By design, the SQED Exceptional Dirichlet condition supports a massive Higgs vacuum in the UV. Compactifying on a circle, the theory flows below the KK scale to a 2d $\cN=(2,2)$ theory of twisted chiral variables, with an effective twisted superpotential. The massive vacuum of the compactified theory is a critical point of this superpotential, or equivalently the endpoint of a BPS gradient flow for the (real part of the) superpotential. 
In other words, the set of points along this flow is precisely a Lefschetz thimble for the twisted superpotential. In the limit $\q\rightarrow 1$, the thimble reduces to the real contour $\Gamma=[0,1]$ for the Veneziano superpotential \eqref{period}.

The appearance of the Exceptional Dirichlet b.c.\ therefore substantiates our supersymmetric proposal: the Coon amplitude should be understood as a deformation of the Veneziano period integral in the UV.
As an application of the SQED Exceptional Dirichlet b.c.\ , we give a novel formula \eqref{vortices} for the Coon amplitude, as a sum of contour integrals over vortex solitons in the massive Higgs vacuum.

\subsection{A meromorphic  Coon amplitude} 
A necessary condition for a meromorphic scattering amplitude to admit a
unitary, local quantum field theory or string-theoretic UV completion is that the residues at each physical pole must factorize into a non-negative sum of partial waves.  For the bare Coon
amplitude however, this property fails~\cite{Figueroa:2022onw,Geiser:2022icl,Cheung:2023adk}, as the residue on a $s$-channel pole is not a polynomial in $t$.\\
A possible  cure for this serious  problem is dressing the bare amplitude by a prefactor:
\begin{equation}\label{modified}
A^{\text{cut}}_\q(\alpha_\q(s),\alpha_\q(t)):= \q^{\alpha_\q(s)\alpha_\q(t) }\,A^{\text{bare}}_\q(\alpha_\q(s),\alpha_\q(t)) \, .
\end{equation}
This ensures the residues are polynomials, and numerical analysis of the expansion in partial waves of the residues has shown positivity \cite{Coon:1972qz,Caron-Huot:2016icg,Figueroa:2022onw, Jepsen:2023sia}.
Yet, the dressing introduces another serious problem. The exponential factor spoils meromorphicity by introducing a logarithmic branch cut\cite{Coon:1972qz,Jepsen:2023sia}, emanating in physical variables from the accumulation point $s_\infty$, which was absent in $A^{\text{bare}}_\q$. Unitarity requires positivity also on the branch cut, but the beautiful paper \cite{Jepsen:2023sia} provided compelling evidence that the dressed amplitude $A_\q^{\rm cut}$ fails to be positive along the branch cut. The physical meaning of this branch cut, apparently needed to rescue positivity, while ultimately dooming unitarity, posed an important puzzle for this amplitude. \\

Here, we show that  introducing a cut along $s\in [s_\infty,\infty]$ is not the only resolution for this issue. We propose the following modification of the Coon amplitude:
\begin{table}
    \centering
    \begin{tabular}{c|c|c|c}
        &  $\displaystyle \frac{\Gamma_\q(-S)\,\Gamma_\q(-T)}{\Gamma_\q(-S-T)}$&  $\displaystyle{\color{blue} \q^{ST}}\frac{\Gamma_\q(-S)\,\Gamma_\q(-T)}{\Gamma_\q(-S-T)}$ &  $\boxed{{\color{blue} K(S,T)}\frac{\Gamma_\q(-S)\,\Gamma_\q(-T)}{\Gamma_\q(-S-T)}}$ \\[.5cm]\hline
        Poles at $S,T\in \bZ^+$  & \checkmark & \checkmark & \checkmark\\
        Local  & \color{red}\xmark & \checkmark & \checkmark\\ 
        Positive& \color{red}\xmark & $\checkmark$& $\checkmark$\\
    Meromorphic&$\checkmark$ & \color{red}\xmark& $\checkmark$\\
           \hline
    \end{tabular}   \caption{Comparison of analytical properties between: the bare Coon amplitude $A^{\rm bare}_\q(S,T) = \frac{\Gamma_\q(-S) \Gamma_\q(-T)}{\Gamma_\q(-S-T)}$, the dressed one usually considered in the literature $A_\q^{\rm cut}(S,T)$, and our new proposal $A^{\theta}_\q(S,T) = K(S,T)\, A_\q^{\rm bare}(S,T)$. We have used the shorthand $S= \alpha_\q(s)$, $T = \alpha_\q(t)$.}
    \label{tab:intro table}
\end{table}
\begin{equation}\label{proposal}
  \boxed{  A^{\theta}_\q(S,T) := K(S,T)\;\frac{\Gamma_\q(-S)\,\Gamma_\q(-T)}{\Gamma_\q(-S-T)}\, , \qquad\;\;\;  K(S,T):=\frac{\theta_\q(-\q^{-S+1/2})\,\theta_\q(-\q^{-T+1/2})}{\theta_\q(-\q^{1/2})\,\theta_\q(-\q^{-S-T+1/2})}}
\end{equation}
where $\theta_\q$ is the elliptic theta function, and  we have expressed the amplitude in the additive variables $S = \frac{\log \sigma}{\log \q}$, $T = \frac{\log \tau}{\log \q}$, just to lighten notation.
The new amplitude we construct, $A^{\theta}_\q(S, T)$, is 
a fully meromorphic S-matrix. Importantly, on its poles at $S,T\in\mathbb{Z}^+$, it has the same residues as those of $A_\q^{\rm cut}$, and therefore it shares all  its desirable analytical properties: our amplitude has polynomial and positive residues, crossing symmetry as well as a natural Regge limit, Gross-Mende behavior and large spin limit. It achieves all of the above without introducing a spurious cut, which previously spoiled unitarity via the $\q^{ST}$ dressing.
This illustrates that giving up single-valuedness of the Coon amplitude is not necessary to achieve positivity at the physical poles. The elliptic prefactor $K(S,T)$ resolves the unwanted cut into a tower of  CDD-like poles in the post-accumulation regime $s>s_\infty$. These elliptic CDD also appear in several examples of fully unitary S-matrices in 2d \cite{Zamolodchikov:1979ba,Castro-Alvaredo:2003qhs}, and are usually hidden in the unphysical kinematic regime. 
Furthermore, in the limits $\q\to 0$ and $\q\to 1$, $A^{\theta}_\q$ reduces to the tree-level field theory and the Veneziano amplitudes, respectively:
\beq
A^{\rm tree}(s,t) \xleftarrow{\q\to 0} (\q -1)A^{\theta}_\q(\alpha_\q(s),\alpha_\q(t)) \xrightarrow{\q\to 1} A^{\rm Ven}(\alpha(s),\alpha(t))\, .
\eeq
Therefore, our proposed Coon amplitude interpolates meromorphically between a tree-level field-theory exchange and the Veneziano amplitude. \\
In order to establish whether this modification corresponds  to a physical scattering amplitude, one would need to verify if the new moving poles are consistent with physical unitarity. The analysis of \cite{Jepsen:2023sia}, as well as ours, suggests that the region $s>s_\infty$ requires a  more careful analysis, as it becomes inaccessible to usual physical kinematics in the $\q\to 1$ limit.\\

In our supersymmetric dictionary, the dressed amplitude with a branch cut, $A_\q^{\rm cut}(S,T)$, arises by turning on a bare mixed Chern-Simons term in the 3d bulk, while our meromorphic proposal, $A^{\theta}_\q(S,T)$, couples instead the bulk theory to a particular 2d $\cN=(0,2)$ theory on the boundary, with the same $\q$-automorphy. 
In the end, our proposal achieves the same on-shell physics as in the literature, but without having to sacrifice global meromorphicity and single-valuedness. 

\subsection{Generalizations}
In this paper, we have limited ourselves to presenting only the simplest example of a much wider correspondence. In this subsection, we present some preliminary results that we plan to explore in \cite{AmbrosinoHaouzi}. We then conclude by discussing  future directions and open questions.
\paragraph{$\q$-deformed closed string amplitude.}
Having discussed (color ordered) $\q$-deformed open string amplitudes, it is natural to wonder whether similar ideas can lead to a $\q$-deformed Virasoro-Shapiro amplitude. Various attempts at constructing a $\q$-deformed closed-string amplitude have failed \cite{Geiser:2022icl,Cheung:2024obl}. 
In this regard, the connection to supersymmetric field theory gives a clear candidate. Our half-index is also a hemisphere partition function on $D^2\times_{\q} S^1$, so
in order to consider the full closed string amplitude, it is then natural to upgrade the half-index to the full index, i.e.\ the partition function on $S^2\times_{\q} S^1$. In the holomorphic block language, this is the well-known fact that $S^2\times_{\q} S^1$ admits a Heegaard decomposition as the union of two solid tori, fused together ``trivially'' at their boundary (after reversing orientation). This is roughly the setup that has been considered in the amplitudes literature so far, but supersymmetry suggests other possibilities: identifying the boundaries of the solid tori using an element of the mapping class group $SL(2,\mathbb{Z})$, yielding the partition function on the squashed sphere $S_b^3$ and its Lens space cousins \cite{Dimofte:2011ju,Dimofte:2011py,Imamura:2013qxa,Dimofte:2014zga}. This associates the closed string amplitude to a bipyramid obtained by gluing two open string amplitudes in a way that is reminiscent of the usual KLT relations.

\paragraph{Higher-points and hypergeometric generalization.}
In this paper, we only consider the 4-punctured disk, but our correspondence suggests a generalization to arbitrary higher points. In general, we find that the $\q$-deformation of a Koba-Nielsen amplitude can be obtained as the half-index of an \textbf{abelian} quiver gauge theory; the number of gauge nodes grows by 1 for each additional puncture. In particular, the 4 point Coon amplitude discussed here is the index of a quiver gauge theory with a single $U(1)$ gauge node. If we now consider the 5-point function, it is naturally associated to the half-index of the quiver:
\vspace{-.5cm}

\begin{equation*}
\includegraphics[clip,width=0.5\textwidth]{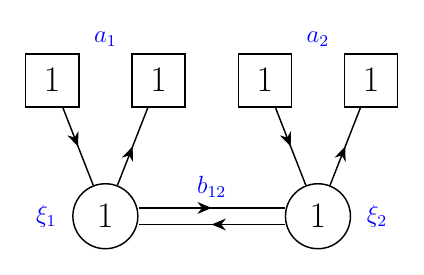}
\end{equation*}

\vspace{-.5cm}
There are now five independent Mandelstam variables, matching nicely the five independent parameters of the gauge theory: two mass splittings $a_1,a_2$, two F.I.\ parameters $\xi_1,\xi_2$, and one bifundamental mass $b_{12}$. Likewise,  $\q$-deformed $N$-points are indices of abelian quivers consisting of $N$ $U(1)$ gauge nodes with additional matter multiplets\footnote{The quivers  generically stop being planar when $N\geq 6$.}. Note that these do not give the Baker-Coon-Romans $N$-point amplitudes \cite{Baker:1970vxk}, that are defined only for $\q>1$,  see also the discussion in \cite{Geiser:2023qqq}, as well as in \cite{Romans:1988qs,Romans:1989di, GonzalezMestres:1975ord}, for other partial attempts to construct these amplitudes. Our correspondence furnishes a natural new candidate for higher-point Coon-generalizations. We plan to address higher-point amplitudes in \cite{AmbrosinoHaouzi}.\\

Another natural generalization is to consider SQED with $2N_f$ chiral multiplets when $N_f>1$. The corresponding half-index will then compute the hypergeometric generalization of the Coon amplitude \cite{Gross:1969db,Pasquetti:2011fj, Rigatos:2023asb,Cheung:2023adk,Wang:2024wcc}.

\subsection{Outlook}
The findings of this paper raise important open questions: 
\begin{itemize}
    \item \underline{When is the index of a given 3d $\cN=2$ theory an amplitude?} \\
    Or formulated differently, what are the necessary conditions on the $3d$ theory side to guarantee positivity, crossing symmetry, and locality of the half index?  
Here, we have limited ourselves to identifying  several cases of supersymmetric indices realizing $\q$-deformed amplitudes. Yet, whether any given 3d $\cN=2$ theory can produce a consistent amplitude is a tantalizing idea that deserves careful investigation. As advertised above, abelian quiver gauge theories with $N_f$ flavors yield candidate for valid amplitudes. It would be then important to explore the case of non-abelian theories to see whether they lead as well to consistent new amplitudes. 
Even more broadly:
can there exist an amplitude when the 3d theory has truly chiral matter? All our working examples so far are \emph{non-chiral} theories, i.e.\ the matter content is always made up of \emph{pairs} of chiral multiplets.  No-go theorems have appeared in similar contexts \cite{Galakhov:2022uyu}. More ambitiously, the positivity of the Coon amplitude depended on turning on a specific mixed Chern-Simons term; what is the corresponding statement for general amplitudes? 
    \item \underline{Is there a physical scattering process realizing the Coon amplitude?}\\
    The Veneziano amplitude was modeled on the physical scattering of mesons and then found to correspond to the physical scattering of open string states. 
 Yet, after many years since the introduction of Coon, we have still failed to find an actual physical model that realizes the properties of this $\q$-deformed amplitude. In \cite{Maldacena:2022ckr}, important progress was made: the authors showed that similar accumulation-of-poles physics happens for rotating strings in $AdS_5$ string backgrounds\footnote{In that proposal, the deformation parameter $\log\q$ plays the role of the $AdS$ curvature.}. However, the precise spectrum was shown to ultimately disagree, and our proposed amplitude similarly fails to match to $AdS_5$ strings. Finding a realization of Coon in some stringy model would shed light on the singularity structure of the post-accumulation regime $s> s_\infty$. 
The connection to open topological strings  we spell out in section \ref{topstrings} is interesting in its own right, and deserves to be explored further.

\item \underline{To cut or not to cut} \\  
\FA{Cfr with:}
In this paper we show that one can preserve single-valuedness of the amplitude, resolving the cut by a tower of CDD poles. This is consistent with the expectation that a tree-level amplitude with poles at the discrete values $s_n= [n]_\q$  should not have a cut, as a cut would mean a continuous spectrum instead. 
Yet ultimately, we do not exclude that a physical realization of the Coon amplitude may force on us the existence of a physical continuum at tree level, meaning a branch cut in the spectrum may be unavoidable. 
After all, this is the case for  the setup of \cite{Maldacena:2022ckr}, corresponding to open string scattering ending on a D-brane in $AdS_5$. In this case, the spectrum coincides with the Coon one, but a cut at large value of $s$ is expected on physical grounds\footnote{We thank Zohar Komargodski and Juan Maldacena for discussions on this point.}. In this regards, the correspondence with  to 3d $\mathcal{N}$  of this paper could of help. 
Besides inserting the dressed factor $\q^{S T}$ in \eqref{modified}, that spoils unitarity \cite{Jepsen:2023sia}, there exist other ways of introducing a branch cut in the spectrum of Coon. For instance, one could start with our meromorphic proposal, and then proceed to smear the boundary degrees of freedom by gauging. Integrating over the boundary kernel along a non-compact contour will typically produce branch cuts, and the hope is that we could do so without violating unitarity.\\

\end{itemize}
\section{Amplitudes from 3d supersymmetry}
\label{3dSUSY}

All fugacities will be complex in this section, and the Coon deformation parameter is chosen so that $|\q|<1$.

\subsection[ 3d \texorpdfstring{$\cN=2$}{N=2} theories \& the half-index]{\boldmath 3d \texorpdfstring{$\cN=2$}{N=2} theories \& the half-index}
We consider 3d $\cN=2$ theories on Minkowski spacetime $\mathbb{R}^{1,1}\times\mathbb{R}_{\leq 0}$, with a 2d boundary on $\mathbb{R}^{1,1}\times\{0\}$ which preserves a 1/2-BPS $\cN=(0,2)$ subalgebra of the 3d $\cN=2$ bulk, as well as a non-anomalous $U(1)_R$ symmetry.

When we talk about a pair of dual theories, we mean that there exist two 3d $\cN=2$ theories ${\cT}$ and $\cT^{\vee}$ defined in the UV, with boundary conditions $\cB$ and $\cB^{\vee}$ respectively, which flow to the same superconformal theory $\cT_{\text{IR}}$ with boundary condition $\cB_{\text{IR}}$ in the IR. In this paper, we take $(\cT,\cB)$ to be 3d SQED, i.e.\ a $U(1)$ gauge theory with 2 chiral multiplets, with Dirichlet b.c.\ on the gauge field and Neumann/Dirichlet b.c.\ for the chiral multiplets. We propose that the dual theory $(\cT^\vee,\cB^\vee)$ is the $\rm X Y Z$ model, i.e.\ three chiral multiplets and a bulk superpotential $W= XYZ$, with Neumann b.c.\ on two of the chiral multiplets and Dirichlet b.c.\ for the remaining chiral: 
\begin{equation}\label{duality}
\begin{tikzcd}[column sep=4.5em, row sep=3.5em]
 (\text{SQED},\{\cD,{\bf N},{\bf D_c}\}) \arrow[rr, leftrightarrow, "{ {\mathrm{Mirror \, \, dual}}}"] \arrow[dr, "\mathrm{Flow}"'] 
&& (\text{XYZ},\{{\bf N},{\bf N},{\bf D}\}) \arrow[dl, "\mathrm{Flow}"] \\
& (\cT_{\text{IR}},\cB_{\text{IR}}) &
\end{tikzcd}
\end{equation}
The duality of the 3d bulk theories is a famous example of 3d $\cN=2$ mirror symmetry, dating back to \cite{Aharony:1997bx,deBoer:1997ka}. More recently, the SQED/XYZ duality was extended to include a nontrivial boundary, and a large number of dual pairs were constructed explicitly \cite{Gadde:2013wq,Okazaki:2013kaa,Dimofte:2017tpi}. Our proposed duality \eqref{duality} should be understood in this context, as a new pair of dual boundary conditions which, to our knowledge, has not been considered in the existing literature. 

Of particular interest to us is an index counting protected BPS operators in the theories $(\cT,\cB)$ and $(\cT^\vee,\cB^\vee)$. Because this index is defined in the UV, it can be evaluated explicitly from the classical fields making up the Lagrangian of the theories. More precisely, the 3d $\cN=2$ theory has four real supercharges, two of which are preserved by the 2d $\cN=(0,2)$ subalgebra; the \emph{half-index} is a character over the vector space of local operators at the boundary, in the cohomology of one of those two supercharges \cite{Gadde:2013wq,Dimofte:2017tpi}:  
\begin{equation}
\label{halfindex}
{\cZ}_{(\cT,\cB)}  = {\rm Tr}\left[(-1)^R\, {q_3}^{J_3+\frac{R}{2}} \;  x^{\ff_{\text{flav}}} \;  \xi^{\ft_{\text{top}}}  \right]\;\; .
\end{equation}
Above, the operator $J_3$ stands for the Cartan generator of the $U(1)_{J_3}$ symmetry rotating the boundary $\mathbb{R}^2$.  
The operator $R$ is a trial $U(1)_R$  R-symmetry generator defined in the UV. The operator $\ff_{\text{flav}}$ stands for the Cartan generator of all flavor symmetries preserved on the boundary; the conjugate fugacities are chiral multiplet masses, collectively denoted by $\log(x)$. Whenever the 3d theory contains a $U(1)$ gauge factor, there is an additional global symmetry $U(1)_{\text{top}}$ associated with a conserved current defined by the dual of the field strength: ${\cal J}=\frac{1}{2\pi}* F$. The operator $\ft_{\text{top}}$ is the generator of this  $U(1)_{\text{top}}$; the conjugate fugacity is the Fayet-Iliopoulos (F.I.) parameter, denoted by $\log{\xi}$. 
In the above conventions, $J_{3}$ has half-integer eigenvalues and $R$ has integer eigenvalues, so the index is in particular a Taylor series in ${q_3}^{1/2}$.

\subsection{The SQED and XYZ models}

The charge assignment of the various fields is\footnote{Our axial and R-symmetry generators are related to those of \cite{Dimofte:2017tpi} by a linear redefinition:
\begin{equation}
\ff^{\text{DGP}}_{a}=2\,\ff^{\text{ours}}_{a}-\ft^{\text{ours}}_{\text{top}}  \, , \qquad\;\;\;\;\; R^{\text{DGP}}=R^{\text{ours}}+\ft^{\text{ours}}_{\text{top}}\, .   
\end{equation}
}:
\[
\begin{array}{c|cc|ccc}
 & \Phi & \widetilde{\Phi} &  X &  Y & Z \\ \hline
U(1)_{\partial} & 1 & -1 & 0 & 0 & 0 \\
U(1)_{a} & 1/2 & 1/2 & 1 & 0 & -1 \\
U(1)_{\text{top}} & 0 & 0 & 0 & 1 & -1 \\
U(1)_R & 0 & 0 & 0 & 0 & 2
\end{array}
\]
On the XYZ side, our charge assignment is compatible with the existence of a superpotential $W=XYZ$ in the bulk. In particular, $W$ has $R$-charge 2 and is neutral under the remaining global symmetries. The ${\bf D}$ b.c.\ on the chiral multiplet $Z$ sets $Z=0$ at the boundary, so $W=XYZ$ vanishes there, and supersymmetry is therefore preserved in the UV. 

On the SQED side, the $\cD$ b.c.\ sets $A_{\mu|\partial}=0$ for the gauge field at the boundary. Since the only gauge transformations that preserve this constraint are constant ones along the boundary, the gauge symmetry is broken to a global (flavor) symmetry $U(1)_{\partial}$ there. The chiral multiplets $\Phi$ and $\widetilde{\Phi}$ have charge +1 and -1 under this $U(1)_{\partial}$, respectively. We give $\Phi$ Neumann boundary conditions ${\bf N}$, and we give $\widetilde{\Phi}$ ``generic'' Dirichlet b.c.\ ${\bf D_c}$ (defined below).  In 3d $\cN=4$ terminology, these SQED b.c.\ go by the name of \textbf{Exceptional Dirichlet} \cite{Bullimore:2016nji}\footnote{In 3d $\cN=4$ theories, the vector multiplet contains an additional $\cN=2$ adjoint chiral multiplet, which would have given ${\bf D}$ b.c.\ and is absent for us. Moreover, the Exceptional Dirichlet b.c.\ depends on a choice of ``polarization'' for the flavor fields: in our SQED setup, this is the freedom to assign either $({\bf N},{\bf D_c})$ or $({\bf D_c},{\bf N})$ b.c.\ to the chiral fields $(\Phi,\widetilde{\Phi})$. As will be clear from \eqref{LHSindex}, the Coon amplitude implicitly picks the $({\bf N},{\bf D_c})$ polarization.}.

Furthermore, we turn on a bare mixed Chern-Simons term at level 1 in the bulk:
\begin{equation}\label{mixedCS}
\frac{1}{2\pi}\int A_{\text{top}}  \wedge dA_{\partial}\, ,
\end{equation}
where $A_{\text{top}}$ and $A_{\partial}$ are background gauge fields for the topological $U(1)_{\text{top}}$ and boundary flavor $U(1)_{\partial}$ symmetries, respectively.

The half-index of the theory $(\text{XYZ},\{{\bf N},{\bf N},{\bf D}\})$ is easy to write down: 
\begin{equation}\label{XYZindex}
{\mathcal Z}^{\text{XYZ}}_{\{{\bf N},{\bf N},{\bf D}\}} =\frac{(a\, \xi\, ; \q)_{\infty}} {(a\, ; \q)_{\infty}\, (\xi\, ; \q)_{\infty}}\, \qquad\qquad (x\, ; \q)_{\infty}:=\prod_{k=0}^{\infty}(1-x\, \q^k)\, .
\end{equation}
This is explained as follows: the three chiral multiplets can be written in terms of $\cN=(0,2)$ chiral and Fermi multiplets, as $(X,\Psi_X)$, $(Y,\Psi_Y)$, and $(Z,\Psi_Z)$. Then, imposing ${\bf N}$ b.c.\ on $X$ and $Y$ sets the $\cN=(0,2)$ Fermi multiplets to zero:
\begin{equation}
\Psi_X|_{\partial} = \Psi_Y|_{\partial} = 0 \, .
\end{equation}
The denominators in \eqref{XYZindex} are the contributions of the surviving boundary scalars (and their n-th derivatives) in the $X$ and $Y$ chiral multiplets. Meanwhile, imposing ${\bf D}$ b.c.\ on $Z$ sets the $\cN=(0,2)$ chiral multiplet to zero:
\beq
Z|_{\partial} = 0 \, .
\eeq
The numerator is the contribution of the surviving boundary fermions (and their n-th derivatives) in the $\Psi_Z$ Fermi multiplet.\\

The index of the theory $(\text{SQED},\{\cD,{\bf N},{\bf D_c}\})$ is more subtle:
\begin{equation}\label{SQEDindex}
{\mathcal Z}^{\text{SQED}}_{\{\cD,{\bf N},{\bf D_c}\}} =\frac{1}{(\q\, ; \q)_{\infty}}\,\sum_{m\in\mathbb{Z}} \xi^m\frac{(\q^{1+m}\, ; \q)_{\infty}} {(\q^m\,a\, ; \q)_{\infty}}\, .   
\end{equation}
The derivation proceeds in two steps: First, we  write down the index of a slightly different theory, $(\text{SQED},\{\cD,{\bf N},{\bf D}\})$,
meaning the chiral multiplet $\widetilde{\Phi}$ is given standard Dirichlet b.c.\ ${\bf D}$.
Even though the $U(1)$ gauge symmetry is broken along the boundary, there is a residual gauge symmetry
orthogonal to it, and associated gauge-invariant operators (and their n-th derivatives) contribute
\beq
\frac{1}{(\q\, ; \q)_{\infty}}
\eeq
to the index. If we explicitly introduce a fugacity $u\in\mathbb{C}^*$ for the generator of the boundary $U(1)_{\partial}$ symmetry, and impose $({\bf N},{\bf D})$ b.c.\ for the chiral multiplets $(\Phi,\widetilde{\Phi})$, the index then reads:
\begin{equation}
\frac{1}{(\q\,  ; \q)_{\infty}}\, \frac{(\q\, u\,a^{-1/2}\, ; \q)_{\infty}} {(u\,a^{1/2}\, ; \q)_{\infty}}\, .   
\end{equation}
Crucially, this expression receives non-perturbative corrections at the boundary coming from BPS monopole operators. The half-index is then a sum over all magnetic fluxes $m\in\mathbb{Z}$ through a hemisphere. Moreover, all electrically charged states acquire spin, meaning the $U(1)_{\partial}$ flavor fugacity $u$ is shifted everywhere to $u\rightarrow \q^m\, u$. After including the contribution of the bare mixed $U(1)_{\text{top}}$-- $U(1)_{\partial}$ Chern-Simons term, the complete formula for the half-index takes the form:
\begin{equation}\label{SQEDindexD}
{\mathcal Z}^{\text{SQED}}_{\{\cD,{\bf N},{\bf D}\}} =\frac{1}{(\q\, ; \q)_{\infty}}\,\sum_{m\in\mathbb{Z}} \xi^m\,\frac{(\q^{1+m}\, u\,a^{-1/2}\, ; \q)_{\infty}} {(\q^m\,u\,a^{1/2}\, ; \q)_{\infty}}\, .   
\end{equation}

The second step is to implement a generic Dirichlet b.c.\ ${\bf D_c}$ for the chiral multiplet $\widetilde{\Phi}$. This is done by initiating an RG flow along the boundary to a new b.c.\ , where $\widetilde{\Phi}$ is set to a constant (nonzero) background chiral superfield ${\bf c}\in\mathbb{C}^*$:
\begin{equation}
{\bf D}\; \text{b.c.\ } :\;\; \widetilde{\Phi}|_{\partial}=0\, , \qquad \longrightarrow \qquad  {\bf D_c}\; \text{b.c.\ } :\;\; \widetilde{\Phi}|_{\partial}={\bf c}\neq 0\, .    
\end{equation}
The b.c.\ ${\bf D_c}$ breaks the boundary flavor symmetry $U(1)_{\partial}$, but the diagonal combination of generators
$\ff_{\text{diag}}:=\ff_{a} + \frac{1}{2}\, \ff_{\partial}$ is preserved under the flow, since $\widetilde{\Phi}$ has charge $(-1,1/2)$ under $U(1)_{\partial}\times U(1)_a$.\\

At the level of the index, then, the ${\bf D_c}$ b.c.\ is preserved only when the weight of $\widetilde{\Phi}$ is set to 1:
\beq\label{Dcconstraint}
u^{-1}\, a^{1/2} = 1\, .
\eeq
We use this constraint to eliminate the fugacity $u$ entirely from the index, by setting $u=a^{1/2}$ inside \eqref{SQEDindexD}. The end result is precisely the index \eqref{SQEDindex}.

\subsection{The Coon amplitude and 3d mirror symmetry}

We propose that the theories $(\text{SQED},\{\cD,{\bf N},{\bf D_c}\})$ and $(\text{XYZ},\{{\bf N},{\bf N},{\bf D}\})$  are mirror dual to each other, where the usual dictionary maps the SQED meson $\Phi \widetilde{\Phi}$ to the chiral field $X$, and the monopole operators of SQED to the chiral fields $Y$ and $Z$. A nontrivial check is to compute the 't Hooft boundary anomalies of each theory; if we denote the field strengths of $U(1)_{\partial},\, U(1)_a,\, U(1)_{\text{top}},\, U(1)_R$ respectively by ${\bf f}, {\bf a}, \boldsymbol{\xi},$ and ${\bf r}$, then the anomaly polynomials are:
\begin{equation}
\mathcal{I}^{\text{XYZ}}_{\{{\bf N},{\bf N},{\bf D}\}}
= - \underbrace{
\frac{1}{2}({\bf a}- {\bf r})^2}_{{\bf N}  \text{ for } X} 
- \underbrace{
\frac{1}{2}(\boldsymbol{\xi}- {\bf r})^2}_{{\bf N}  \text{ for } Y} 
+ \underbrace{
\frac{1}{2}(-\boldsymbol{\xi}- {\bf a}+ {\bf r})^2}_{{\bf D}  \text{ for } Z} =  -\frac{1}{2}{\bf r}^2+ {\bf a}\, \boldsymbol{\xi}\, ,
\end{equation}
and
\begin{equation}
\begin{aligned}
\mathcal{I}^{\text{SQED}}_{\{\cD,{\bf N},{\bf D_c}\}}
&=
\underbrace{
 2 \,{\bf f}  \,{\boldsymbol{\xi}}
}_{\text{bulk C.S.}}
- \underbrace{
\frac{1}{2}{\bf r}^2
}_{\cD\text{ for gauge}}
- \underbrace{\frac{1}{2}({\bf f}+\frac{{\bf a}}{2} - {\bf r})^2}_{{\bf N}  \text{ for } \Phi}
+\underbrace{\frac{1}{2}(-{\bf f}+\frac{{\bf a}}{2} - {\bf r})^2}_{{\bf D}  \text{ for } \widetilde{\Phi}}\left.\rule{0pt}{4.5ex}\right|_{{\bf f}=\frac{{\bf a}}{2}}\\
&=
-\frac{1}{2}{\bf r}^2+ {\bf a}\, \boldsymbol{\xi}-\frac{1}{2}{\bf a}^2 + {\bf a}\, {\bf r}\, .
\end{aligned}
\end{equation}
Above, the condition ${\bf f}=\frac{{\bf a}}{2}$ implements the generic Dirichlet b.c.\ ${\bf D_c}$; it is the boundary anomaly counterpart to the condition \eqref{Dcconstraint} used in the index.
We see that the anomalies agree up to $U(1)_a$-- $U(1)_a$ and $U(1)_a$-- $U(1)_R$ contact terms in SQED: these anomalies do not affect the half-index, and can be removed by hand by further turning on associated bare Chern-Simons terms.\\
More nontrivially, the indices ${\mathcal Z}^{\text{SQED}}_{\{\cD,{\bf N},{\bf D_c}\}}$ and ${\mathcal Z}^{\text{XYZ}}_{\{{\bf N},{\bf N},{\bf D}\}}$ agree exactly:
\begin{equation}\label{equalindices}
\boxed{\;\;\underbrace{\frac{1}{(\q\, ; \q)_{\infty}}\,\sum_{m\in\mathbb{Z}} \xi^m\frac{(\q^{1+m}\, ; \q)_{\infty}} {(\q^m\,a\, ; \q)_{\infty}}}_{{\displaystyle\mathcal Z}^{\text{SQED}}_{\{\cD,{\bf N},{\bf D_c}\}}} = \underbrace{\frac{(a\, \xi\, ; \q)_{\infty}} {(a\, ; \q)_{\infty}\, (\xi\, ; \q)_{\infty}}}_{{\displaystyle\mathcal Z}^{\text{XYZ}}_{\{{\bf N},{\bf N},{\bf D}\}}}\;\;}
\end{equation}
This can be proved analytically by invoking the $\q$-binomial identity, for instance.\\
To make contact with the Coon amplitude, we propose that the F.I.\ parameter and axial mass should be interpreted as Mandelstam variables
\beq\label{qMandelstam}
\xi := \q^{-S} \, , \qquad a:= \q^{-T}\, .
\eeq
Then, notice that the left-hand side of the mirror identity \eqref{equalindices} can be massaged into:
\begin{align}
 \frac{1}{(\q\, ; \q)_{\infty}}\,\sum_{m\in\mathbb{Z}} \xi^m\frac{(\q^{1+m}\, ; \q)_{\infty}} {(\q^m\,a\, ; \q)_{\infty}} &=  \frac{1}{(\q\, ; \q)_{\infty}}\,\sum_{m=0}^{\infty} \xi^m\frac{(\q^{1+m}\, ; \q)_{\infty}} {(\q^m\,a\, ; \q)_{\infty}}\nonumber\\
 &=  \frac{1}{(\q\, ; \q)_{\infty}}\,\sum_{m=0}^{\infty} \q^m \q^{-m(S +1)}\frac{(\q^{1+m}\, ; \q)_{\infty}} {(\q^m\,\q^{-T}\, ; \q)_{\infty}}\nonumber\\
 &= \frac{1}{(\q\, ; \q)_{\infty}\,(1-\q)}\int_0^1 d_\q z\; z^{-S-1}\, \frac{(\q\,z\,;\q)_{\infty}}{(\q^{-T} \, z\,;\q)_{\infty}}\label{LHSindex}\, .
 \end{align}
In the first line, we used that $(\q^{1+m}\, ; \q)_{\infty}=0$ if $m<0$, so the monopole sum is really a sum over positive charges only. In the second line, we made the Mandelstam variable substitution \eqref{qMandelstam}. In the third line, we invoked the definition of the Thomae-Jackson $\q$-integral \cite{thomae1869beitrage,jackson1910qdefinite}:
\begin{equation}
\int_0^1 d_\q z\; f(z) := (1-\q)\sum_{m=0}^{\infty} \q^m\, f(\q^m)\, ,
\end{equation}
which is a $\q$-analog of the usual Riemann integral (it reduces to it in the limit $\q\rightarrow 1$).

Meanwhile, the right-hand side of the mirror identity \eqref{equalindices} is a simple ratio of $\q$-Gamma functions \cite{heine1847untersuchungen,thomae1869beitrage,jackson1904generalization} in Mandelstam arguments:
\begin{equation}\label{RHSindex}
\frac{(a\, \xi\, ; \q)_{\infty}} {(a\, ; \q)_{\infty}\, (\xi\, ; \q)_{\infty}}=\frac{1}{(\q\, ; \q)_{\infty}\,(1-\q)}\,\frac{\Gamma_\q(-S)\,\Gamma_\q(-T)}{\Gamma_\q(-S-T)}\, ,
\end{equation}
where 
\begin{equation}
\Gamma_\q(x):=(1-\q)^{1-x}\, \frac{(\q\,;\q)_{\infty}}{(\q^x \, ;\q)_{\infty}}\, , \qquad\qquad |\q|<1\,,\;\;\;\; x\in\mathbb{C}\backslash\{0,-1,-2, \, \ldots\}\, .
\end{equation}
Comparing \eqref{LHSindex} with \eqref{RHSindex}, we obtain a well-known identity:
\begin{equation}
\int_0^1 d_\q z\; z^{-S-1}\, \frac{(\q\,z\,;\q)_{\infty}}{(\q^{-T} \, z\,;\q)_{\infty}} = \frac{\Gamma_\q(-S)\,\Gamma_\q(-T)}{\Gamma_\q(-S-T)}\, .
\end{equation}
This is nothing but the $\q$-integral definition of the $\q$-Beta function, which is precisely the 1-parameter generalization of the Veneziano relation between an integral and a ratio of Gamma functions. Indeed, in the original Mandelstam variables, the identity reads: 
\begin{equation}
\begin{aligned}
A^{\text{bare}}_\q(\alpha_\q(s),\alpha_\q(t)):=\frac{\Gamma_\q(-\alpha_\q(s))\,\Gamma_\q(-\alpha_\q(t))}{\Gamma_\q(-\alpha_\q(s)-\alpha_\q(t))}=\int_0^1 d_\q z\; z^{-\alpha_\q(s)-1}\, &{\color{blue}\frac{(\q\,z\,;\q)_{\infty}}{(\q^{-\alpha_\q(t)} \, z\,;\q)_{\infty}}} \, ,\\[0.6em]
&\;\;\;\;\;\;\;\Big\downarrow \q \to 1 \\[0.1em]
   A^{\rm Ven}(\alpha(s),\alpha(t)) := \frac{\Gamma(-\alpha(s))\,\Gamma(-\alpha(t))}{\Gamma(-\alpha(s)-\alpha(t))} =
    \int_0^1 dz \;z^{-\alpha(s)-1}\, &{\color{blue}(1-z)^{-\alpha(t)-1} }\, .
    \end{aligned}
\end{equation}
We have just shown that the evaluation of the $\q$-Beta integral follows from the equality of 3d indices in mirror SQED and XYZ theories!\\

So far, we have only described how the \emph{bare} Coon amplitude $A^{\text{bare}}_\q(S,T)$ is realized via 3d supersymmetry. We now argue that the modified Coon amplitude $A^{\text{cut}}_\q(S,T)$ \eqref{modified} from the literature, as well as our meromorphic proposal $A^{\theta}_\q(S,T)$ \eqref{proposal}, also have a very natural supersymmetric interpretation.
\subsection{Crossing symmetry of the index}
The $(\text{XYZ},\{{\bf N},{\bf N},{\bf D}\})$ half-index \eqref{XYZindex} is obviously invariant under exchange of the F.I.\ parameter and the axial mass. This immediately implies that the index of the mirror SQED theory can instead be expanded as a monopole sum in the $S$-channel, as:
\begin{equation}\label{equalindices2}
\underbrace{\frac{(a\, \xi\, ; \q)_{\infty}} {(a\, ; \q)_{\infty}\, (\xi\, ; \q)_{\infty}}}_{{\displaystyle\mathcal Z}^{\text{XYZ}}_{\{{\bf N},{\bf N},{\bf D}\}}}=\underbrace{\frac{1}{(\q\, ; \q)_{\infty}}\,\sum_{m\geq 0} \xi^m\frac{(\q^{1+m}\, ; \q)_{\infty}} {(\q^m\,a\, ; \q)_{\infty}}}_{\displaystyle\text{$T$-channel}:\;\, {\mathcal Z}^{\text{SQED}}_{\{\cD,{\bf N},{\bf D_c}\}}} =\underbrace{\frac{1}{(\q\, ; \q)_{\infty}}\,\sum_{m\geq 0} a^m\frac{(\q^{1+m}\, ; \q)_{\infty}} {(\q^m\,\xi\, ; \q)_{\infty}}}_{\displaystyle\text{$S$-channel}:\;\,{\mathcal Z}^{\text{SQED}^{\vee}}_{\{\cD,{\bf N},{\bf D_c}\}}} 
\end{equation}
On the right-hand side, we introduced the theory  $(\text{SQED}^{\vee},\{\cD,{\bf N},{\bf D_c}\})$, whose chiral multiplets $(\Phi^{\vee},\widetilde{\Phi}^{\vee})$ are obtained from the original multiplets $(\Phi,\widetilde{\Phi})$ by swapping the topological $U(1)_{\text{top}}$ and axial $U(1)_a$ symmetries.\\
After multiplying both expansions by the Coon normalization $(\q\, ; \q)_{\infty}\,(1-\q)$,  the last equality is nothing but the $\q$-analog of the usual crossing symmetry residue expansion of the Veneziano amplitude. While the crossing symmetry is manifest on the XYZ side, it is certainly not obvious on the mirror SQED side.

\subsection{A contour integral representation of the amplitude via vortices}

In the absence of axial mass, it is well known that the 3d SQED gauge theory admits a 1-dimensional Higgs branch $\mathbb{C}$, whose gauge-invariant modulus is provided by the meson field $X=\Phi\,\widetilde{\Phi}$. After turning on the axial mass, this continuous branch is lifted to a single massive vacuum, which happens to be precisely compatible with the Exceptional Dirichlet b.c.\ , and further admits BPS vortex solitons.\\

Our SQED half-index is known to count these vortices, as a certain generating function of Witten indices for a 1d gauged quantum mechanics on their worldline. The inquisitive reader is encouraged to consult Appendix~\ref{contour} for details. Here, we only report the final result; for positive real values of the 3d F.I.\ and axial mass parameters,
\begin{equation}\label{vortices}
\boxed{A^{\text{bare}}_\q(\xi,a)=\frac{(1-\q)(\q;\q)_{\infty}}{(a;\q)_{\infty}}\sum_{m=0}^{\infty}\xi^m\,\frac{\mathfrak n_m}{m!}
\oint
\prod_{i=1}^{m}
\frac{\dd\phi_i}{2\pi i\,\phi_i}
\,
\frac{
\prod_{i\neq j}
\left(1-\frac{\phi_i}{\phi_j}\right)
}{
\prod_{i,j}
\left(1-\q\frac{\phi_i}{\phi_j}\right)
}
\prod_{i=1}^{m}
\frac{1-a\phi_i}{1-\phi_i}\, ,}
\end{equation}
where
\begin{equation}
\mathfrak n_m
=
(-1)^{m}\, 
\q^{\frac{m(m-1)}{2}}\, ,
\end{equation}
and the contour is defined to enclose the poles at:
\begin{equation}
\phi_i=\q^{i-1} \, .
\end{equation}
Note that we previously understood the summation label $m$ as the monopole charge of operators on
the SQED Coulomb branch. The same integer $m$ is now interpreted as the vortex charge of operators on the Higgs branch instead.
\subsection{Dressing the Coon amplitude via supersymmetry}\label{ssec:dressingSUSY}
The reason the bare Coon amplitude needs an improvement in the first place is that its residue at physical poles is only a Laurent polynomial instead of an ordinary polynomial in the physical Mandelstam variables:  
\begin{equation}
    \begin{aligned}\label{automorphy}
&A^{\text{bare}}_\q(s,t)\;\text{$s$-channel:\;\; poles are at $S=n\in \mathbb{Z}^+$, \;\; residues are in $\q^{-n T}\cdot\mathbb{C}[[\q^T]]$}\\
&A^{\text{bare}}_\q(s,t)\;\text{$t$-channel:\;\; poles are at $T=n\in \mathbb{Z}^+$, \;\; residues are in $\q^{-n S}\cdot\mathbb{C}[[\q^S]]$}
    \end{aligned}
    \end{equation}
Namely, the bare Coon half-index does not have correct \emph{$\q$-automorphy} to represent a unitary amplitude. In the supersymmetric language, the $\q$-automorphy of the half-index is its transformation under large shifts 
\begin{equation}\label{qauto}
\xi \,\rightarrow \,\q\, \xi \, , \qquad a\, \rightarrow \, \q\, a\, .
\end{equation}
Mathematically, the half-index is not simply a function of the F.I.\ parameter $\xi$ and the axial mass $a$; it is a meromorphic section of a line bundle over the elliptic torus
\begin{equation}
    \left(\mathbb{C}^*_{\xi}/\q^{\mathbb{Z}}\right) \times \left(\mathbb{C}^*_{a}/\q^{\mathbb{Z}}\right)\, .
\end{equation}
So instead of being invariant under \eqref{qauto}, the index transforms via transition functions on this line bundle:
\begin{equation}
\begin{split}
{\mathcal Z}^{\text{XYZ}}_{\{{\bf N},{\bf N},{\bf D}\}}(\q\, \xi\, , \, a) &= M_{\xi}( \xi\, , \, a)\cdot {\mathcal Z}^{\text{XYZ}}_{\{{\bf N},{\bf N},{\bf D}\}}(\xi\, , \, a)\, \\
{\mathcal Z}^{\text{XYZ}}_{\{{\bf N},{\bf N},{\bf D}\}}(\xi\, , \q\, a) &= M_{a}(\xi\, ,\, a)\cdot {\mathcal Z}^{\text{XYZ}}_{\{{\bf N},{\bf N},{\bf D}\}}(\xi\, , \, a)\, .
\end{split}
\end{equation}
Physically, this transition function data is entirely encoded in the boundary anomalies and contact terms of the theory. Therefore, the lack of unitarity of the bare Coon amplitude can a priori be cured by shifting the anomaly of the theory. There are essentially two ways to achieve this in 3d.\\

A first option is to turn on bare Chern-Simons terms in the bulk, which in turn triggers a boundary 't Hooft anomaly through inflow. 
This is precisely the strategy used in the literature\cite{Figueroa:2022onw,Jepsen:2023sia}: the modified amplitude $A^{\text{cut}}_\q(s,t)$ \eqref{modified}  introduces a new factor
\begin{equation}\label{CSfix}
\q^{ST}=e^{\frac{\log{\xi}\,\log{a}}{\log{\q}}}\, 
\end{equation}
in the XYZ index, which corresponds to turning on a bare Chern-Simons at level 1 for the mixed $U(1)_{\text{top}}$-- $U(1)_a$ symmetry. This in turn adds a term $+2{\bf a}\, \boldsymbol{\xi}$ to the boundary 't Hooft anomaly polynomial, which cures the ``bad $\q$-automorphy'' of the residues \eqref{automorphy}.\\

A second option is to couple the 3d theory to a 2d $\cN=(0,2)$ theory on the boundary\footnote{More generally, one can consider an interface between two 3d theories, with index
\begin{equation}
{\mathcal Z}^{\text{right}}(y)=\sum_{m}\int dx \;K(x,y;m)\,{\mathcal Z}^{\text{left}}(x;m) \, ,
\end{equation}
where the kernel $K$ modifies the $\q$-automorphy through its own anomaly. In this paper, we only need a very special  interface, between the XYZ theory and a trivial theory, i.e.\ a boundary condition. The extra degrees of freedom living on that interface make up our 2d boundary theory.
The kernel then reduces to multiplication by the elliptic genus of that boundary theory.}. This is the path we follow in this paper: our proposed amplitude $A^{\theta}_\q(S,T)$ in \eqref{proposal}  introduces the factor
\begin{equation}\label{thetafix}
\frac{\theta_\q(-\q^{1/2}\,\xi)\,\theta_\q(-\q^{1/2}\,a)}{\theta_\q(-\q^{1/2})\,\theta_\q(-\q^{1/2}\,\xi\, a)}\, ,\qquad\qquad\;\; \theta_\q(x):=(x\,;\q)_{\infty}\, (\q\,x^{-1}\,;\q)_{\infty}\, ,
\end{equation}
which is the index contribution of two free chiral multiplets $C_{\xi\, a}$, $C_1$, and two free Fermi multiplets $\Gamma_{\xi}$, $\Gamma_{a}$. In the XYZ theory, the charges of these multiplets are summarized in the Table below\footnote{Alternatively, one could work in the SQED mirror frame, where the 2d multiplets would be further charged under the $U(1)_{\partial}$ boundary group before imposing the generic Dirichlet b.c.\ .}:
\[
\begin{array}{c|cccc}
 & \Gamma_{\xi} & \Gamma_{a} &  C_{\xi\, a} & C_1 \\ \hline
U(1)_{a} & 0 & 1 &  1 & 0 \\
U(1)_{\text{top}} & 1 &  0 & 1 & 0 \\
U(1)_R & 1 & 1 & 1 & 1
\end{array}
\]
These multiplets contribute a term 
\begin{equation}
+ \underbrace{
(\boldsymbol{\xi}+ {\bf r})^2}_{\Gamma_{\xi}} 
+ \underbrace{
({\bf a} + {\bf r})^2}_{\Gamma_{a}} 
- \underbrace{
(\boldsymbol{\xi}+ {\bf a}+ {\bf r})^2}_{C_{\xi\, a} }- \underbrace{({\bf r})^2}_{C_{1} } =  -2\,{\bf a}\, \boldsymbol{\xi} 
\end{equation}
to the boundary 't Hooft anomaly polynomial, which once again cures the residues \eqref{automorphy}; details are given in the next Section\footnote{In writing the anomaly polynomials, we are using the conventions of \cite{Dimofte:2017tpi}. Then, the expressions \eqref{CSfix} and \eqref{thetafix} have opposite anomaly polynomials, but the same $\q$-automorphy.}.

As a last remark, it may seem surprising to introduce a neutral chiral multiplet $C_1$, which contributes only a pure $U(1)_R$-- $U(1)_R$ contact term to the anomaly, and the denominator $\theta_\q(-\q^{1/2})^{-1}$ to the index. The reason we include this multiplet explicitly is to ensure that $A^{\rm bare}_\q$ and $A^{\theta}_\q$ share the same normalization: the theta function ratio is $1$  for $\zeta \neq \q^\bZ$, which in particular holds for our choice of trial R-charges.\\

\section{A meromorphic \& positive Coon amplitude}
\label{Anew modification}

The supersymmetric realization of the $\q$-deformed Beta integral as a half-index suggests that a natural $\q$ deformation of the Coon amplitude is:
\beq
\boxed{
A^{\theta}_\q(S,T) := \frac{\theta_\q(-\q^{-S+1/2})\,\theta_\q(-\q^{-T+1/2})}{\theta_\q(-\q^{1/2})\,\theta_\q(-\q^{-S-T+1/2})}\;\frac{\Gamma_\q(-S)\,\Gamma_\q(-T)}{\Gamma_\q(-S-T)}}
\eeq
Where the variables $S,T$ are convenient additive shorthand for the logarithmic Coon trajectories:
\beq\begin{aligned}\label{traj}
\sigma(s)&=1+(s-m^2)(\q-1),
\qquad &\tau(t)&=1+(t-m^2)(\q-1), \\
S&=\alpha_\q(s)=\frac{\log\sigma(s)}{\log \q},
\qquad &T&=\alpha_\q(t)=\frac{\log\tau(t)}{\log \q}.
\end{aligned}
\eeq
In this section we take $\q$ in the range $0<\q<1$\footnote{From the SUSY perspective, one could take $|\q|< 1$, yet taking complex values of $\q$ would spoil the reality of the amplitude on the real axis, so here we restrict to the real slice thereof.}.  This amplitude has all the following properties: 
\begin{itemize}
\item Crossing-symmetry.
    \item Discrete  spectrum  with logarithmic trajectories accumulating at $s_\infty$
    \beq 
 s_n=m^2+\frac{1-\q^n}{1-\q}=m^2+[n]_\q,
  \qquad
  [n]_\q:=\frac{1-\q^n}{1-\q}, \qquad   s_\infty
  =
  m^2+\frac{1}{1-\q}.
\eeq
and analogously in the $t$-channel.
\item The residue on its poles $s_n$ is a polynomial in $t$ of degree $n$ satisfying positivity constraints.
\item It is a meromorphic amplitude: the logarithmic cut is resolved into a series of poles, corresponding to  elliptic/CDD images outside of the physical domain in the Mandelstam plane.
\item It interpolates \textbf{meromorphically} between a tree-level scattering amplitude at $\q \to 0$ and the Veneziano amplitude at $\q \to 1$.
\end{itemize}

\paragraph{Meromorphic completion.}
It is worth noting that any amplitude of the form
\begin{equation} 
A^{\theta}_\q(S,T|\zeta) = K(S,T|\zeta)\;A^{\text{bare}}_\q(S,T), \qquad K(S,T|\zeta)= \frac{\theta_\q(\zeta\,\q^{-S})\,\theta_\q(\zeta\,\q^{-T})}{\theta_\q(\zeta)\,\theta_\q(\zeta\,\q^{-S-T})}
\end{equation}
shares all the above features, provided that: 
\beq 
\zeta \neq \q^\bZ \, ,
\eeq
corresponding to points where $\theta_\q(\zeta)$ vanishes, up to normalizing factors. This is therefore a genuine freedom of this meromorphic extension. In Section~\ref{ssec:dressingSUSY}, we made the choice of setting the UV trial R-charges to 1 for all 2d multiplets,
\beq 
\boxed{A^{\theta}_\q(S,T) := A^{\theta}_\q(S,T|-\q^{1/2})}\, ,
\eeq
so that the amplitude reduces to a field theory amplitude as $\q\to 0$, in standard normalization. This choice also happens to be the fixed point of the involution
\beq 
z\mapsto \frac{\q}{z}, \qquad\theta_\q(z)=-z^{-1}\,\theta_\q(z^{-1})\, .
\eeq
$K(S,T|\zeta)$ is an elliptic meromorphic completion of the factor $\q^{ST}$ as it has the important feature of having
the same finite-shift automorphy as $\q^{ST}$. Indeed, using 
\beq
\theta_\q(\q^n z)=(-1)^n \q^{-n(n-1)/2}z^{-n}\,\theta_\q(z),
\qquad n\in \bZ ,
\eeq
one finds, for $n\in\bZ^+$ (and analogously in $t$ by crossing symmetry of $K(S,T|\zeta)$),
\beq
K(S= n,T|\zeta)=\q^{nT}.
\eeq
Thus, on the pole lattice of the $\q$-Beta
function, the theta kernel reproduces precisely the value of the usual
Gaussian factor \(\q^{st}\). This ensures that: 
\beq 
\boxed{\Res{S = n} A^{\theta}_{\q}(S,T) = \Res{S = n} A^{\rm cut}_{\q}(S,T)}
\eeq
Since the residues of $A_\q(s,t|\zeta)$ agree exactly with those of the
standard exponentially modified Coon amplitude\cite{Figueroa:2022onw}, all residue-level partial-wave positivity properties of the modified Coon
amplitude are inherited unchanged by the theta-completed amplitude\cite{Figueroa:2022onw}:
\beq
\begin{split}\Res{S=n} A^{\theta}_\q(S,T)
&=
\frac{1-\q}{\log \q}\,
\frac{(\q^{T+1};\q)_n}{(\q;\q)_n}, \quad {\rm or \:equivalently   }\quad 
\Res{s=s_n}A^{\theta}_\q(s,t)=-\,\q^n\frac{(\q\,\tau(t);\q)_n}{(\q;\q)_n}.
\end{split}
\eeq
That is, it is a polynomial of degree $n$ in $t$ with positive Gegenbauer expansion for the same range as in \cite{Figueroa:2022onw}. 
Thus, the upgrade: 
\beq 
{\color{blue} \q^{ST}} \,\frac{\Gamma_\q(-S)\,\Gamma_\q(-T)}{\Gamma_\q(-S-T)} \to {\color{blue}K(S,T|\zeta) } \, \frac{\Gamma_\q(-S)\,\Gamma_\q(-T)}{\Gamma_\q(-S-T)}
\eeq
is a genuine uplift of the exponential factor to a 
meromorphic function of the multiplicative variables.  Indeed, although the standard exponential factor $\q^{ST}$ is harmless-looking in the additive variables \(s,t\), it becomes multivalued alone the Regge trajectories \eqref{traj}:
\beq
\sigma(s) = \q^{-S},\quad \tau(t)=\q^{-T}, \qquad \q^{ST}
=
\exp\left(\frac{\log\sigma\,\log\tau}{\log \q}\right)
\eeq
which depends on the choice of branches of $\log\sigma(t)$ and $\log\tau(t)$. The theta kernel instead can be written directly as
\beq
K(\sigma(s),\tau(t)|\zeta)
=
\frac{\theta_\q(\zeta\sigma(s))\theta_\q(\zeta\tau(t))}
{\theta_\q(\zeta)\theta_\q(\zeta\sigma(s)\tau(t))} ,
\eeq
and is  single-valued and meromorphic in \(\sigma,\tau\). In this
sense, the theta factor is a meromorphic completion of the logarithmic
cocycle \(\q^{ST}\). Note that also the deformed Beta is manifestly single-valued: 
\beq 
\frac{\Gamma_\q(-S)\Gamma_\q(-T)}{\Gamma_\q(-S-T)}
=
\frac{(1-\q)(\q;\q)_\infty(\sigma\tau;\q)_\infty}
{(\sigma;\q)_\infty(\tau;\q)_\infty}.
\eeq
\paragraph{New poles.}
\begin{figure}
    \centering
\includegraphics[width =\linewidth]{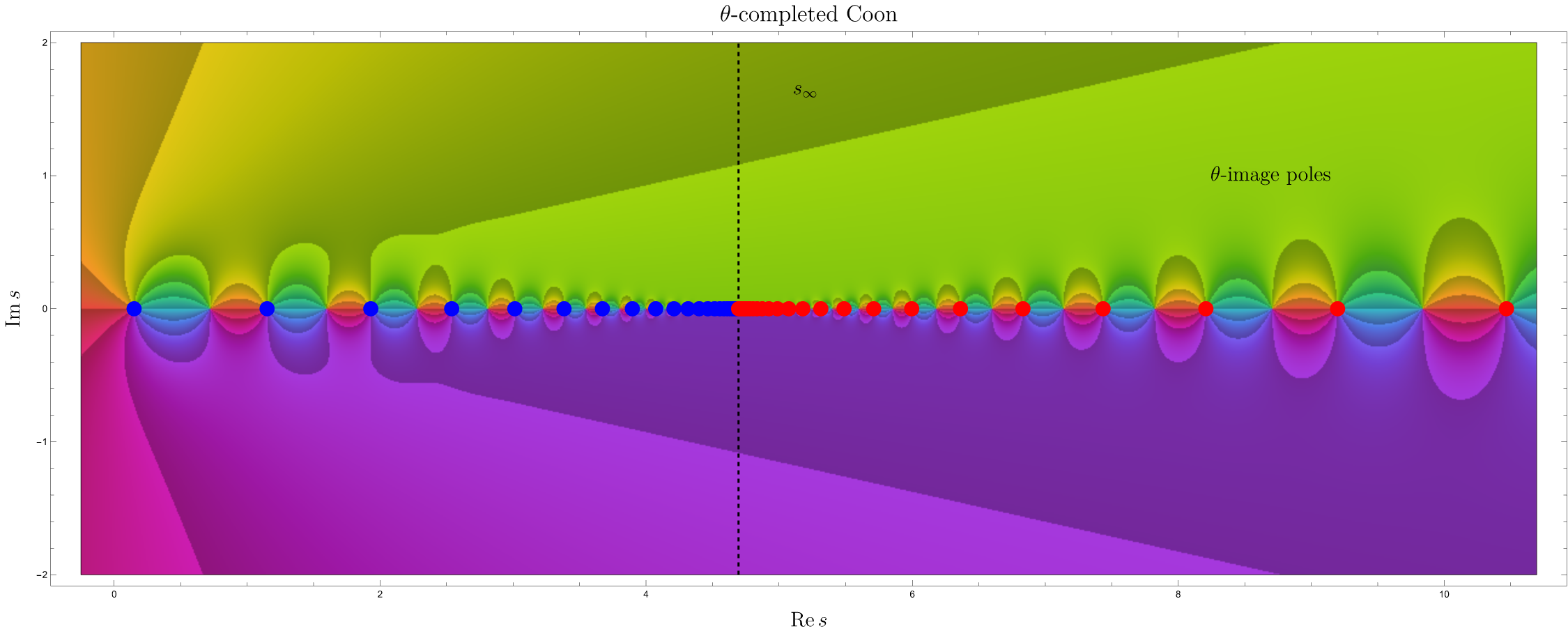}
    \caption{Plot of $A_\q(\alpha_\q(s),\alpha_\q(t))$ for a typical slice at $-\infty<t< t_\infty$. In $\color{blue} \text{blue}$, the physical poles at $s=m^2+[n]_\q$; in $\color{red} \text{red}$, the poles due to the meromorphic $\theta$ completion.}
    \label{fig:thetaCoon}
\end{figure}
The meromorphic completion of $\q^{ST}$ to $K(S,T|\zeta)$ resolves the branch cut emanating from $s_\infty$ into a tower of unphysical (moving) poles. These poles occur at the divisor where the denominator
theta-function vanishes:
\beq\label{newpoles}
  \theta_\q(\zeta \q^{-S-T})=0 \Leftrightarrow  \zeta \q^{-S-T}=\q^N 
\eeq
For the distinguished half-period choice $\zeta=-\q^{1/2}$, this gives
singularities on the logarithmic cover at:
\beq
  S+T
  =
  \frac12-N+\frac{i\pi(2k+1)}{\log \q},
  \qquad N,k\in\mathbb Z .
\eeq
Thus, for fixed real $T$, the image poles do not lie on the real $S$-axis and vice versa.
They are shifted away from the physical trajectory sheet by
\beq |{\rm Im} \,  S |=\frac{\pi(2k+1)}{|\log \q|} \eeq
Along the Regge trajectories,
  $\sigma(s)=\q^{-S},\; \tau(t)=\q^{-T}$,
the same divisor 
  $\sigma\tau=-\q^{1/2-N}$ does not have any image  in the below-accumulation chamber
  $\sigma,\tau>0$. 
  For $\tau(t)>0$, the extra poles in the physical $s$-plane all lie  beyond the accumulation point
  at $s>s_\infty$ ($\sigma <0$). Note that this includes unphysical values  $0<t<t_\infty$.  The analyticity  region  ${\rm Re}\, s< s_\infty,\; {\rm Re}\, t<t_\infty $, or equivalently $|{\rm Im }S|,|{\rm Im }T| < \frac{\pi}{\log \q} $, is completely free of $\theta$-function poles. 
These poles are not ordinary \(s\)-channel
particles (and indeed their location depends on $t$). They could be  viewed as
CDD-like elliptic image poles due to meromorphic continuation. In 2d elliptic integrable models\cite{Zamolodchikov:1979ba,Castro-Alvaredo:2003qhs}, these are needed to have fully crossing symmetric and unitary S-matrices. Indeed, there are some similarities with the structure of these examples; the difference is the greatly simplified single-channel kinematics of the 2d amplitude. \\
In our case, 
the new poles
resolve  the branch-cut ambiguity of the logarithmic factor \(\q^{ST}\) by
a single-valued meromorphic divisor, while leaving the physical Coon pole
residues unchanged.
They correspond to extra meromorphic
data, rather than new finite-level exchanged
states  \footnote{Another supporting idea for this interpretation is the ambiguity of these new poles exactly as in the case of the CDD-ones: their location is a function of the $\zeta$ parameter that is effectively free. }.  

One of the resolutions of the negativity of the branch cut offered in \cite{Jepsen:2023sia} was to deem the post-accumulation chamber unphysical in $\q$-kinematics. What our discussion entails is indeed that the presence of a branch cut is not unavoidable to obtain a positive amplitude in the analytic region $s< s_\infty,\; t< t_\infty$. 
Our modification is just as good as the one in the literature in this region, but manages to not spoil single-valuedness. 
It would be important to provide a physical model that exhibits this feature, otherwise the physical significance remains unclear, as well as their admissibility. 

\paragraph{\boldmath $\q \to 1$ and $\q \to 0$  limits.}
Another crucial feature of our modification is that it interpolates meromorphically between the tree-level exchange at $\q \to 0$: 
\beq 
\lim_{\q \to 0^+}(\q-1) A^{\theta}_\q(\alpha_\q(s),\alpha_\q(t)) = \frac{1}{s-m^2} + \frac{1}{t-m^2} -1 \, ,
\eeq
and the Veneziano amplitude at $\q \to 1$:
\beq 
\lim_{\q \to 1^-}(\q-1) A^{\theta}_\q(\alpha_\q(s),\alpha_\q(t)) = A^{\rm Ven}(\alpha(s),\alpha(t)) = \frac{\Gamma(-s-m^2)\Gamma(-t-m^2)}{\Gamma(-s-t-2m^2)} \, .
\eeq
These two limits crucially follow from the following two identities precisely engineered by the $\zeta$ factors in $K(S,T|\zeta)$\footnote{The limit $\q\to 0$ is $\zeta$-independent at order $\mathcal{O}(1)$ provided that $\zeta = c\, \q^\delta$ for some constants $c,\delta$.}:
\beq 
\begin{split}
K(S,T|\zeta)&\xrightarrow{\q\to 1} \frac{(-\zeta)^{-S}(-\zeta)^{-T}}{(-\zeta)^{-S-T}} + \mathcal{O}(\q -1)= 1 + \mathcal{O}(\q -1)\, ,\\
K(S,T|\zeta)&\xrightarrow{\q\to 0} 
\frac{
(1-\zeta/\sigma)(1-\zeta/\tau)
}{
(1-\zeta)(1-\zeta/(\sigma\tau))
}+\mathcal{O}(\q)  ={\Bigg|}_{\zeta=c\, \q^\delta} 1 +\mathcal{O}(\q)\, .
\end{split}\eeq
In the $\q\to 1$ limit, the Coon trajectory linearizes again as $s_\infty \to \infty$, and the extra poles are pushed away, while in the $\q\to 0$ limit, all the physical and unphysical poles but the one at $s,t=m^2$ collapse at the point $s_\infty\xrightarrow{\q\to 0} m^2+1$, becoming smooth there. 

Hence, compared to $A^{\rm cut}_{\q}(S,T)$, where the amplitude was multivalued for $\q \neq 0,1,$ but single-valued at the limit points, our amplitude has the nice feature of meromorphically interpolating between tree-level scattering with a single pole and the Veneziano amplitude with a full Regge trajectory.

\subsection[Topological string expansion]{\boldmath Topological string expansion}\label{topstrings}
It is well-known that the 3d half-index we have been considering has an interpretation as an open topological string partition function, in the topological A-model \cite{Dimofte:2010tz,Beem:2012mb}.
To study the kinematic limit, the following $\q$-expansion of  $K(S,T)$ is useful:
\beq
K(S,T) = \q^{ST}
\left[\frac{\sin(\pi(S+T))}
{\sin(\pi S)\sin(\pi T)}+
4\sin(2\pi(S+T))\widetilde{\q}^{1/2}
+
O(e^{-4\pi^2/g_s})
\right]\,,\;\;\;\;\; \tilde{\q}=e^{4\pi^{2}/\log \q}
\eeq
where the expansion in $\tilde{\q}=e^{4\pi^{2}/\log \q} = e^{-4\pi^{2}/|\log \q|}$ is bounded for $S,T\in \bR$.  \\
An analogous expansion for the $\q$-Beta gives:
\begin{align}
\nonumber\frac{\Gamma_\q(-S)\Gamma_\q(-T)}{\Gamma_\q(-S-T)} = -\mathcal{C}_\q\, \q^{-ST}&\Big[
1+ \tilde{\q} \big(3+2\sin(\pi S)\sin(\pi T)\cos(\pi (S+T))\big) +\mathcal{O}(\tilde\q^2)
\Big]\\
\times
&\underbrace{\exp\bigg(-\sum_m \frac{\q^m}{m(1-\q^m)}(\q^{mS}+\q^{mT} - \q^{m(S+T)})\bigg)}_{\displaystyle\mathcal{Z}_{\rm open}}\label{qGamamexp}
\end{align}
where (anticipating the notation in \eqref{gsdef}):
\beq 
C_\q =  \frac{1}{2}(1-\q)\sqrt{\frac{g_s}{2\pi}}(\q;\q)^2_\infty  (\q/\tilde{\q})^{1/8}
\eeq
This expansion as an open topological string partition function is well known from the index point of view  \cite{Dimofte:2010tz,Beem:2012mb}. 
Indeed,  identifying: 
\beq \label{gsdef}
\q = e^{-g_s}, \qquad \tilde{\q} = e^{-4\pi^2/g_s}
\eeq
the expansion \eqref{qGamamexp} is the open topological $A$-model string partition function for a Lagrangian brane at position $q^{S+1/2}$ of the external leg of a resolved conifold with $\mathbb{P}^1$ K\"ahler parameter $\tau$. The two  terms $$\sum_{m\geq 1 } \frac{q^{m(S+1/2)} - q^{m(S+T+1/2)}}{m[m]}$$ are the standard external brane contributions in the resolved conifold, with K\"ahler size $T$\footnote{We thank Suriyah Rajalingam Kannagi for extended discussions on this. }.

\begin{figure}
    \centering
    \begin{subfigure}{0.48\textwidth}
        \centering
        \includegraphics[height=5cm]{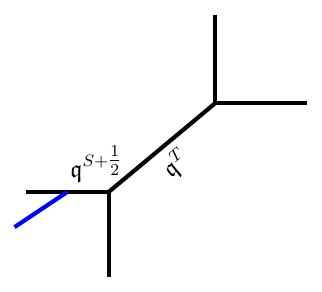}
        \caption{Topological A-model}
        \label{fig:toric}
    \end{subfigure}
    \hfill
    \begin{subfigure}{0.48\textwidth}
        \centering
    \includegraphics[height=5cm]{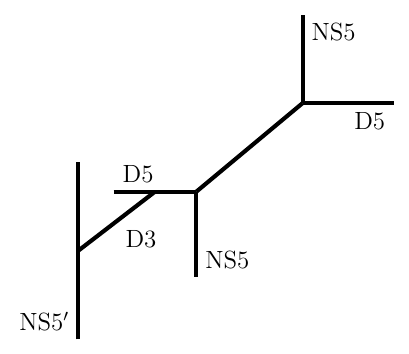}
        \caption{Type IIB brane web}
        \label{fig:IIB}
    \end{subfigure}
    \caption{Figure \ref{fig:toric}: Brane configuration in the topological A-model that computes $\mathcal{Z}_{\rm open}$ in \eqref{topstringfull}; in {\color{blue} blue}, the Lagrangian brane at position $\q^{S+1/2}$. Figure \ref{fig:IIB}: type IIB  $(p,q)$ brane web engineering the  3d $\mathcal{N}=2$ SQED theory living on the D3 brane.}
    \label{fig:branes}
\end{figure}
The remaining term is due to the relative normalization of the closed amplitude $Z_{\rm cl}(e^{-t})= \sum_{m\geq 1} \frac{e^{-mt}}{m[m]^2} $ with and without an extra brane in the geometry that increases the flux through the $\mathbb{P}^1$ by one unit of $g_s$.  Indeed, this term coincides with the ratio $Z_{\rm cl}(e^{-t})/Z_{\rm cl}(e^{-t+g_s}) = \sum_{m\geq 1} \frac{e^{-mt}}{m[m]}$. 
The toric geometry is shown in Figure~\ref{fig:toric}.

The perturbative term in $g_s$, captured by the topological string expansion, is dressed by a non-perturbative series in $g_s$ controlled by $e^{-4\pi^2/g_s}$.
Putting the two together we find: 
\beq \label{topstringfull}
A_\q(S,T) =-\mathcal{C}_\q \, \mathcal{Z}_{\rm open}(S,T) \, \left( \frac{\sin(\pi (S+T))}{\sin(\pi S)\sin(\pi T)} + 4\sin(2\pi(S+T)) \, e^{-2\pi^2/g_s}+\mathcal{O}(e^{-4\pi^2/g_s}) \right) 
\eeq
Note that the new poles \eqref{newpoles}  are non-perturbative in this picture and come from resummation of the non-perturbative terms in $g_s$.
\paragraph{Regge \& Gross-Mende.}
From the expansion above \eqref{topstringfull}, one can easily extract the Regge and Gross-Mende (fixed angle) limit  of $A_\q(S,T)$:
\begin{align}
 {\rm (Regge):}&\qquad   S\to+\infty \quad (\Leftrightarrow s\to s_\infty),
  \qquad
 &T \quad  \text{fixed}\\
 ({\rm Gross\text{--}Mende}):&\qquad S,T\to+\infty, \qquad  &S/T\quad {\rm fixed} 
\end{align}
The amplitude is bounded  in the limits  and modulated by the trigonometric function in \eqref{topstringfull}.

\paragraph{The index as a topological string. }
One way to explain why our 3d half-index is an open topological string partition function is via M-theory/Type IIB duality \cite{Dimofte:2010tz}. The starting point is the following (local) background in M-theory:
\begin{equation*}
\begin{aligned}
&\text{M-theory}: & S^1\times \mathbb{C}_z\times \mathbb{C}_w \times X\;\;\;\;\, \\
&  & \cup\qquad\qquad\, \cup\;\;\;\;\, \\
&\text{M5-brane}:  & S^1\times D^2\times \{.\} \times {\color{blue} L_{\rm blue}} \\
&  & \qquad\qquad\, \;\;\;\;\, \\
&\text{M2-branes}:  & S^1\times \{.\}\times \{.\} \times \Sigma\;\;\;\,\, \\
\end{aligned}
\end{equation*}
where $X$ is the resolved conifold, ${\color{blue} L_{\rm blue}}\subset X$ is the Lagrangian brane (a $T^2_{\rm toric}$-fibration over the blue line in Figure~\ref{fig:toric}),  $D^2\subset \mathbb{C}_z$ is a finite disk, and $\Sigma\subset X$ is a  holomorphic disk ending on the M5-brane, $\partial\Sigma\subset{\color{blue} L_{\rm blue}}$. The M-theory background is nontrivially fibered: going around $S^1$ once rotates $\mathbb{C}_z\times \mathbb{C}_w$:
\begin{equation}
\label{actionM}
(\,z\,,w\,)\;\,\mapsto\; (\,\q\, z\,,\,\q^{-1}\,w\,) \, .
\end{equation}
The first fact we need is that the open topological string partition function on $X$ is counting the above M2-branes ending on M5-branes in this M-theory background \cite{Gopakumar:1998ii,Aganagic:2000gs}.

Next, if we view the toric Calabi–Yau $X$ as a $T^2_{\rm toric}$ fibration, we can reinterpret M-theory on $T^2_{\rm toric}$ as Type IIB string theory on a (different) circle \cite{Leung:1997tw}. The toric skeleton, which records the degeneration loci of the $(p,q)$ cycles for the $T^2_{\rm toric}$, becomes a web of  $(p,q)$ 5-branes in the IIB frame. The M5-brane turns into a D3-brane, and the open M2-branes turn into  $(p,q)$ strings stretching between the D3-brane and a $(p,q)$ 5-brane.

Then, it is well-known that the low energy theory on the worldvolume of the D3 brane is 3d $\cN=2$ SQED. The gauge coupling of SQED is kept finite by introducing an additional NS5$'$-brane ``out of the page'' for the D3-brane to end on, so that it has finite extent in the base direction of ${\color{blue} L_{\rm blue}}$\cite{Hanany:1996ie,Elitzur:1997fh,Hanany:1997vm,Kitao:1998mf,Hanany:2003hp}; see Figure~\ref{fig:IIB}.

The count of M2-branes becomes the Euclidean partition function of the D3-brane theory in Type IIB, defined on
\begin{equation}\label{Omegaback}
D^2\times_\q S^1 \, ,
\end{equation}
where $D^2$ is the disk (or equivalently a hemisphere),  and $S^1$ plays the role of the Euclidean time circle. The notation ``$\times_\q$'' indicates that the background is nontrivially fibered, as was previously the case in M-theory: going around $S^1$ once rotates the disk as
\beq\label{action}
z\mapsto \q\, z\, .
\eeq
The partition function is defined as a trace over the Hilbert space of states on $D^2$, and computed via equivariant localization, by quantizing the zero modes of strings on the D3-brane\footnote{The fugacity $\log(\q)$ is still conjugate to the Cartan generator $J_3$ of the rotation symmetry $U(1)_{J_3}$, now understood as an eigenvalue for azimuthal rotations on the hemisphere. The partition function is then interpreted as the generating function of operators which saturate the BPS bound $D-R-J_3 \geq 0$, where $D$ is the Cartan generator of the $U(1)$ rotation along the circle $S^1$.}. 
In fact, this partition function on Euclidean space is known to coincide with the 3d half-index \eqref{halfindex} in Minkowski space\footnote{Note that the boundary of the background \eqref{Omegaback} is a torus $T^2$. This naturally explains why computing pure 2d $\cN=(2,0)$ boundary contributions to the 3d half-index gives theta functions, i.e. the elliptic genus of the boundary $T^2$ theory. This fact is obscured in the Minkowski presentation.}, up to 1-loop contributions of boundary mixed 't Hooft anomalies \cite{Yoshida:2014ssa,Bullimore:2020jdq}. We have thus achieved our goal of identifying the 3d half-index as an open topological string partition function.\\

Finally, note that the NS5 and NS5$'$-branes are offset from each other, which forces the D3-brane to end on a D5-brane in order to preserve supersymmetry. This turns on the real F.I. parameter in the low energy SQED theory on the D3-brane, which in turn locks the theory into a massive Higgs vacuum in the IR. The Exceptional Dirichlet b.c. we imposed in SQED is compatible with this Higgs phase, as we review in Appendix \ref{contour} below, and since it is defined in the UV, it is natural to ask if it can be realized via additional branes in Type IIB. A good candidate preserving 2d $\cN=(2,0)$ supersymmetry would be the D5$''$-brane of type B, in the nomenclature of  \cite{Chung:2016pgt}. We defer this matter for further investigation.

\paragraph{Acknowledgments.}
We are indebted to Davide Gaiotto for extensive discussions. We also thank him, Zohar Komargodski and  Andrea Guerrieri for discussions and comments on an earlier draft of this paper. We also thank   Kevin Costello, Gabriel Cuomo, Matthias Gabardiel, Jaume Gomis, Suriyah Rajalingam Kannagi, Juan Maldacena, Shu-Heng Shao, Pedro Vieira, and Yifan Wang, for various  discussions and suggestions on different aspects of this work.
Research at Perimeter Institute is supported in part by the
Government of Canada through the Department of Innovation, Science and Economic Development Canada and
by the Province of Ontario through the Ministry of Colleges and Universities.
\appendix

\section*{Appendix}
\section{Details on the contour integral representation of the amplitude}
\label{contour}

In this Appendix, we use supersymmetry to explain why the Coon amplitude admits the contour integral representation \eqref{vortices}.

\subsection{Counting vortices on the Higgs branch}

We start by reviewing briefly how the Euclidean partition function of 3d $\cN=2$ SQED on $D^2\times_{\q} S^1$ from \eqref{Omegaback} counts BPS vortices on the \emph{Higgs branch}.

In the absence of axial mass, it is well known that the 3d SQED gauge theory admits a 1-dimensional Higgs branch $\mathbb{C}$, whose gauge-invariant modulus is provided by the meson field $X=\Phi\,\widetilde{\Phi}$. After turning on the axial mass, this continuous branch is lifted to a single massive vacuum, which can be determined from the (real) supersymmetric vacuum equations:
\begin{equation}\label{SUSYequations}
|\widetilde{\Phi}|^2 - |{\Phi}|^2 = \zeta^{\bR}_{3d} \, ,\;\; \qquad
\left(\sigma + \frac{m^{\bR}_a}{2}\right)\Phi = 0\, , \;\;\qquad \left(-\sigma + \frac{m^{\bR}_a}{2}\right)\widetilde{\Phi} = 0\, ,
\end{equation}
where $\sigma$ is the real scalar field of the 3d $\cN=2$ vector multiplet, and $\zeta^{\bR}_{3d}$ and $m^{\bR}_a$ are the real F.I.\ and axial mass parameters, respectively\footnote{In the index, we used exponentiated parameters $\xi$ and $a$ instead; they are related to $\zeta^{\bR}_{3d}$ and $m^{\bR}_a$ via
\begin{equation}
\log{|\xi|}:=-\zeta^{3d}_{\rm F.I.}\, , \qquad\;\; \log{|a|}:=-m^{\bR}_a\, .
\end{equation}
In particular, the sign condition $\zeta^{\bR}_{3d}>0$ translates to $|\xi|<1$ in our index notation. The phases of the parameters $\xi$ and $a$ are provided by the holonomies of background gauge fields along the circle \eqref{Omegaback}.}.  For positive F.I.\ parameter $\zeta^{\bR}_{3d}>0$, the Higgs vacuum solution to \eqref{SUSYequations} is reached by letting $\widetilde{\Phi}$ condense:
\beq\label{vacsol}
|\widetilde{\Phi}|^2=\zeta^{\bR}_{3d} \, , \;\; \qquad {\Phi} = 0\, , \;\;\qquad \sigma=\frac{m^{\bR}_a}{2}\, .
\eeq
Meanwhile, the field ${\Phi}$ acquires a mass $m^{\bR}_a$.
Crucially, our choice of b.c.\ $(\text{SQED},\{\cD,{\bf N},{\bf D_c}\})$ is compatible with this vacuum: the generic Dirichlet b.c.\ on $\widetilde{\Phi}$ precisely sets $\sigma|_{\partial}=m^{\bR}_a/2$ at the boundary (recall \eqref{Dcconstraint}), in agreement with the above vacuum solution. The Neumann b.c.\ on $\Phi$ simply leaves its boundary scalar unconstrained, allowing it to fluctuate around its 0 vev with mass $m^{\bR}_a$.\\

The Higgs solution provided by the $\widetilde{\Phi}$ condensate admits BPS vortex solitons; for positive F.I.\ parameter $\zeta^{\bR}_{3d}>0$, such vortices are solutions to the equations:
\begin{equation}
\begin{aligned}
D_{\bar z}\widetilde{\Phi}&=0\, ,\\
*\,F&=\zeta^{\bR}_{3d}-|\widetilde{\Phi}|^2\, ,
\end{aligned}
\end{equation}
where the vortex number is the magnetic flux
\beq
m:=-\frac{1}{2\pi}\int_{D^2} F\in\mathbb{Z}^+\, .
\eeq
This integer $m$ is a winding number for the phase of $\widetilde{\Phi}$: at the edge of the disk, where
\beq
\widetilde{\Phi}\sim \sqrt{\zeta^{\bR}_{3d}}\,e^{i\varphi(\theta)}\, ,
\eeq
a vortex number $m$ implies
\beq
\varphi(\theta+2\pi)-\varphi(\theta)=2\pi m\, .
\eeq
Since a continuous complex field with nonzero phase winding cannot remain nonzero everywhere on the disk, $\widetilde{\Phi}$ must vanish, and its zeros are the positions of the vortices.
More precisely, the first equation says that the field $\widetilde{\Phi}$ is a holomorphic section of a degree-$m$ line bundle over the disk, hence $\widetilde{\Phi}$ has $m$ zeros. In a given gauge, this configuration of $m$ vortices at positions $z=z_\alpha$ reads:
\begin{equation}\label{vortexpositions}
\widetilde{\Phi}(z)= \sqrt{\zeta^{\bR}_{3d}}\,\prod_{\alpha=1}^m (z-z_\alpha)=\sqrt{\zeta^{\bR}_{3d}}\;(z^m+s_1\, z^{m-1} + s_2\, z^{m-2}+ \ldots + s_m)\, .
\end{equation}
For $m$ abelian vortices, the \emph{moduli space of vortices} is famously the $m$-th symmetric product of $\mathbb{C}$ \cite{jaffe1980vortices}:
\begin{equation}
\mathcal{M}^{\mathrm{vortex}}_m = \text{Sym}^m(\mathbb{C})\, , \qquad\;\;\; \text{dim}_{\bC}(\mathcal{M}^{\mathrm{vortex}}_m)=m \, ,
\end{equation}
so the coefficients $s_1, \ldots, s_m$, which are elementary symmetric functions of the positions $z_\alpha$, are good collective coordinates on this moduli space.\\

The partition function of SQED on $D^2\times_\q S^1$ is a refined count of these vortices, performed using equivariant localization at the fixed point $z=0$ on the disk under the rotation action \eqref{action}. The contributing vortex configurations are therefore living on loops wrapping \(S^1\) and sitting at the origin of $D^2$. Since each coordinate $s_i$ in \eqref{vortexpositions} carries weight $\q^i$ under this action, an $m$-vortex configuration is counted via an equivariant character of the vector space of holomorphic functions on $\mathcal{M}^{\mathrm{vortex}}_m$:
\begin{equation}
\text{Ch}_{\q}\bC[s_1,\ldots,s_m] = \prod_{i=1}^{m}\frac{1}{1-\q^i}:=\frac{1}{(\q;\q)_{m}}\, .
\end{equation}
In words, each holomorphic coordinate from the condensing field $\widetilde{\Phi}$ contributes a bosonic oscillator  upon quantization, resulting in a total of $m$ $\q$-Pochhammers in the denominator.

There is an additional matter bundle over the vortex moduli space $\mathcal{M}^{\mathrm{vortex}}_m$, which arises from the fluctuations of the massive $\Phi$ field, and contributes $\q$-Pochhammers in the numerator:
\begin{equation}
{(a;\q)_{m}}\, .
\end{equation}
Altogether, the vortex partition function is a generating series
\begin{equation}\label{vortexpoch}
\sum_{m\geq 0}\xi^m \, {Z}_m^{\mathrm{vortex}}\, , \qquad\;\;\;\;{Z}_m^{\mathrm{vortex}}:=\frac{(a;\q)_m}{(\q;\q)_m}
\end{equation}
over all vortex numbers.\\

An alternate (and quick!) way to derive this vortex count is to start from our SQED index. There, we had previously understood the integer $m$ as the monopole charge of operators on the SQED Coulomb branch. Now, the same integer $m$ can instead be interpreted as the vortex charge of operators on the Higgs branch. Explicitly, using the $\q$-Pochhammer identity
\begin{equation}
(z;\q)_m  = \frac{(z;\q)_\infty}  {(\q^m \,z;\q)_\infty}\, , \qquad\;\;\; m\in\mathbb{Z}^+ \, ,
\end{equation}
we immediately find:
\begin{align}
{\mathcal Z}^{\text{SQED}}_{\{\cD,{\bf N},{\bf D_c}\}}
&=
\frac{1}{(\q;\q)_\infty}
\sum_{m\ge 0}
\xi^m
\frac{(\q^{1+m};\q)_\infty}{(\q^m a;\q)_\infty}
=
\frac{1}{(a;\q)_\infty}
\sum_{m\geq 0}
\xi^m
\frac{(a;\q)_m}{(\q;\q)_m}:={Z}^{\mathrm{pert}}\;
\sum_{m\geq 0}\xi^m \, {Z}_m^{\mathrm{vortex}}\, ,
\end{align}
where 
\begin{equation}\label{pertpoch}
{Z}^{\mathrm{pert}}=\frac{1}{(a;\q)_\infty}
\end{equation}
stands for a perturbative 1-loop determinant, present already at vortex number $m=0$, while ${Z}_m^{\mathrm{vortex}}$ is the vortex contribution \eqref{vortexpoch}.
\subsection{Contour integral from a 1d gauged quantum mechanics}

In the adiabatic approximation, the low-energy dynamics of the above vortices is governed by a 1d $\cN=2$ effective supersymmetric quantum mechanics along the circle $S^1$ in the Euclidean picture\footnote{By 1d $\cN = 2$ supersymmetry, we mean the reduction of 2d $\cN = (0,2)$ supersymmetry to 1 dimension.}. In practice, this is achieved by promoting the vortex positions to vary \emph{slowly} as a function of time,  $z_\alpha \rightarrow z_\alpha(t)$, with characteristic time scale
\begin{equation}
\tau_{\text{motion}} \;\gg\;  \frac{1}{e_{3d}\, \sqrt{\zeta^{\bR}_{3d}}} \,\, , \;\; \frac{1}{m^{\bR}_a}\, ,
\end{equation} 
relative to the 3d bulk mass scales, so  that the only light degrees of freedom are the vortex collective coordinates. Then, the IR quantum mechanics is a sigma model whose target space is precisely the vortex moduli space of SQED, $\mathcal{M}^{\mathrm{vortex}}_m = \text{Sym}^m(\mathbb{C})$.\\
This quantum mechanics can also be defined in the UV, as a 1d gauged linear sigma model:
\begin{equation*}\label{2quivers}
\includegraphics[clip,width=0.8\textwidth]{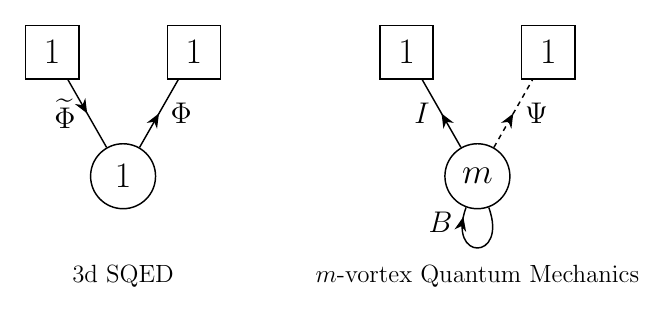}
\end{equation*}

At vortex number $m$, the quantum mechanics has the following $\cN=2$ multiplet content:
\begin{itemize}
    \item $V$: vector multiplet for a $U(m)_{\text{gauge}}$ gauge group,
    \item $B$: adjoint chiral multiplet describing vortex position fluctuations,
    \item $I$: fundamental chiral multiplet from the $\widetilde{\Phi}$ condensate,
    \item $\Psi$: fundamental fermi multiplet from the $\Phi$ field, denoted by a dashed arrow.
\end{itemize}
The charges of the matter multiplets under the various symmetry groups are summarized in the table below:
\[
\begin{array}{c|ccc}
 &  B & I & \Psi \\ \hline
U(m)_{\text{gauge}} & \text{adj} & m & m  \\
U(1)_{a} & 0 & 0 & 1  \\
U(1)_{J_3} & 1 & 0 & 0 \\
U(1)_R & 0 & 0 & 0
\end{array}
\]
The partition function of the vortex theory on $S^1$ is the Witten index of the quantum mechanics, computed as the cohomology of the same supercharge used in the 3d half-index\footnote{The Witten index has literally the same expression as in 3d, expression \eqref{halfindex}, but the trace is now taken over the Hilbert space of the 1d quantum mechanics in a fixed $m$-vortex sector.}. For the above charge assignment, the 1-loop determinants evaluate to:
\begin{equation}
Z_m^{\mathrm{vortex}}
=
\frac{\mathfrak n_m}{m!}\;
\overbrace{\oint\;
\prod_{i=1}^{m}
\frac{d\phi_i}{2\pi i\,\phi_i}
\,
\frac{
\prod_{i\neq j}\left(1-{\phi_i}/{\phi_j}\right)
}{
 \underbrace{\prod_{i,j}\left(1-\q\,{\phi_i}/{\phi_j}\right)
}_{\text{adjoint chiral } B}}}^{\text{vector } V}
\;\;\overbrace{\underbrace{\prod_{i=1}^{m}
\frac{1-a\,\phi_i}{1-\phi_i}}_{\text{fund. chiral } I}}^{\text{fund. fermi } \Psi}.
\end{equation}
The constant in front of the integral is:
\begin{equation}
\mathfrak n_m
=
(-1)^{m}
\q^{\frac{m(m-1)}{2}}.
\end{equation}
We have specified the integrand, but one still needs to define the contour, which is well-known to be in one-to-one correspondence with a choice of sign for the real 1d F.I.\ parameter $\zeta^{\bR}_{1d}$ \cite{Cordova:2014oxa,Hori:2014tda,Hwang:2014uwa}. Indeed, when $\zeta^{\bR}_{1d}$ changes sign and crosses
the value $\zeta^{\bR}_{1d}=0$, the Witten index typically jumps due to wall crossing. Here, the real D-term equation
\begin{equation}
  [B,B^\dagger] + I\, I^{\dagger} = \zeta^{\bR}_{1d} \; \mathbb{I}_{m\times m} 
\end{equation}
has no solution when $\zeta^{\bR}_{1d}<0$, so the negative F.I.\ chamber is empty. Meanwhile, when $\zeta^{\bR}_{1d}>0$, the Witten index is nonzero, and the contour prescription goes by the name of Jeffrey-Kirwan residue\footnote{Alternatively, in this rather simple theory, one can introduce an ``$i\epsilon$'' prescription to the integrand and work with unit circle contours on $\mathbb{C}^*$.}. The inquisitive reader is invited to consult the references above for details. For our purposes, it suffices to say that the poles enclosed by the contours are at:
\beq
\phi_i=\q^{i-1}\, .
\eeq
It is not hard to show that the corresponding residue evaluates to:
\begin{equation}
Z_m^{\mathrm{vortex}}
=
\prod_{i=0}^{m-1}
\frac{1-a\q^i}{1-\q^{i+1}}
=
\frac{(a;\q)_m}{(\q;\q)_m}\, ,
\end{equation}
so we recover the 3d expression \eqref{vortexpoch}. This concludes our derivation of the contour formula \eqref{vortices} for Coon.

\section{Amplitudes and enumerative geometry}

Although the mirror identities appearing in this paper are new to amplitude physics, they are all rudimentary instances of more sophisticated statements known in enumerative geometry.  
Indeed, the $\q$-Beta function (what we physicists call the Coon amplitude) is an elementary example of a \emph{vertex function}  \cite{Okounkov:2015spn}, also called a K-theoretic I-function. An early version of this statement was formulated already in the 1990s \cite{Matsuo:1993}.

In detail, the Higgs branch moduli space of 3d $\cN=2$ SQED is a K\"ahler GIT quotient by the complexified gauge group\footnote{One subtlety here is that ${\mathcal X}$ is \emph{not} a holomorphic symplectic quotient in this paper. In physics, we say that the 3d theory only has $\cN=2$ supersymmetry instead of $\cN=4$. A vertex function usually depends on a choice of \emph{polarization}, which  is a choice of holomorphic Lagrangian splitting
of the tangent bundle of ${\mathcal X}$; such a definition is not available in the $\cN=2$ case, but a weaker notion of polarization still makes sense for the $\q$-Beta function \cite{Haouzi:2023doo}.},
\begin{equation}
{\mathcal X}=\left(\mathbb{C}_{\Phi}\times\mathbb{C}_{\widetilde{\Phi}}\right) /\!\!/_{_{\zeta^{\bR}_{3d}>0}}\,\mathbb{C}^* \simeq \mathbb{C}_{\Phi\widetilde{\Phi}} \, .
\end{equation}
The moduli space of vortices from the previous Appendix is known mathematically as a moduli space of stable based quasimaps \cite{ciocan2014stable}:
\begin{equation}
f:\, \mathbb{P}^1 \rightarrow {\mathcal X} \, .
\end{equation}
``Stable'' means that the map $f$ can take unstable values in ${\mathcal X}$ only at finitely many points, the locations of the vortex cores. ``Based'' is a condition imposed at $\infty\in\mathbb{P}^1$: the $U(1)$ gauge field is required to be flat there, and the matter fields $\Phi$ and $\widetilde{\Phi}$ are required to settle asymptotically into the isolated Higgs vacuum \eqref{vacsol} there. If we denote this vacuum by the label ${\tilde p}$, and the degree of the quasimap by the vortex number $m$, then the moduli space of degree-$m$ quasimaps in the Higgs vacuum ${\tilde p}$ is commonly written as
\begin{equation}
\text{QM}_{\tilde p}({\mathcal X};m) \simeq \text{Sym}^m(\mathbb{C})\, .
\end{equation}
Then, the $\q$-Beta function is a generating function of equivariant Euler characteristics of the virtual structure sheaf over QM$_{\tilde p}({\mathcal X};m)$, up to normalization by constant maps:
\begin{equation}\label{quasimap}
\begin{aligned}
A^{\text{bare}}_\q(\xi,a) &=
\frac{(1-\q)(\q;\q)_{\infty}}{(a;\q)_{\infty}}
\;\sum_{m\geq 0}\;\xi^m\;
\chi_{_T}\left(\text{QM}_{\tilde p}({\mathcal X};m),
{\mathcal O}_{vir}\right)\\
&= \frac{(1-\q)(\q;\q)_{\infty}}{(a;\q)_{\infty}}\;\sum_{m\geq 0}\;\xi^m\;\frac{(a;\q)_m}{(\q;\q)_m}\, .
\end{aligned}
\end{equation}
Above, the F.I. parameter $\xi$ is the K\"{a}hler parameter keeping track of the map degree, and $a$ is the axial mass; see the quiver on page~\pageref{quiverSQED}. The count is performed in equivariant K-theory with respect to the torus action ${\mathbb{C}^*_{\q}\times\mathbb{C}^*_a}$:
\begin{equation}
z\in\mathbb{P}^1 \; \mapsto \q\, z \; , \qquad\;\;\; \Phi\,\widetilde{\Phi}\in {\mathcal X} \; \mapsto a\; \Phi\,\widetilde{\Phi} \; .
\end{equation}
The virtual structure sheaf ${\mathcal O}_{vir}$ is the tensor product of: 
\begin{itemize}
    \item The structure sheaf of the $m$-vortex moduli space, whose tangent space contributes $(\q;\q)^{-1}_m$ to the Euler characteristic. This denominator comes from the bosonic zero modes of the condensing field $\widetilde{\Phi}$.
    \item The obstruction matter bundle over the $m$-vortex moduli space, whose K-theoretic Euler class contributes $(a;\q)_m$. This numerator comes from fermionic zero modes of the field $\Phi$.
\end{itemize}
To conclude, we highlight various statements from our paper which can readily be translated to enumerative geometry. We will not attempt to be thorough or rigorous; instead, we refer the interested reader to the recent set of lecture notes by Okounkov \cite{Okounkov33Lectures} for details and references.
\begin{itemize}
\item 
The dressed amplitude \eqref{modified} changes the framing of the vertex function via the factor $e^{\log\xi\, \log a/\log\q}$; this is a section of a mixed K\"{a}hler/flavor line bundle, which requires choosing a branch of the logarithm. Our theta function proposal \eqref{proposal} is instead the elliptic Thom class of a virtual bundle, which we engineered to be isomorphic to the former K\"{a}hler/flavor line bundle. In particular, our proposed amplitude is a meromorphic representative section of that same line bundle.

\item In our conventions, the K\"ahler parameter $\xi$ satisfies $|\xi|<1$, and the $\q$-Beta vertex function \eqref{quasimap} is naturally analytic in that chamber. The crossing symmetry identity \eqref{equalindices2} is the statement that the $\q$-Beta function admits another presentation as a different vertex function, where the roles of the K\"ahler parameter $\xi$ and mass parameter $a$ are exchanged. In particular, the new vertex function is now analytic in the chamber $|a|<1$ instead. This phenomenon is a consequence of the elliptic stable envelope construction \cite{Aganagic:2016jmx,Aganagic:2017smx,Dinkins:2020xvn}, which relates vertex functions in different chambers in this precise fashion. The $\q$-Beta function is elementary enough that the envelopes are just constant here. In order to see nontrivial envelopes, one needs to study $\q$-hypergeometric generalizations of the Coon amplitude.

\item The identity \eqref{qGamamexp} is the statement that the $\q$-Beta function can be interpreted as a certain refined count of open DT invariants, where the underlying noncompact Calabi-Yau 3-fold geometry and Lagrangian brane are displayed in Figure~\ref{fig:toric}.

\end{itemize}

\newpage
\begin{spacing}{0.70}
{\small \bibliography{coon.bib}}
\end{spacing}

\end{document}